# Developing Global Aerosol Models based on the Analysis of 30-Year Ground Measurements by AERONET (AEROEX Models) and Implication on Satellite based Aerosol Retrievals.

Manoj K Mishra[1], Shameela S F[1,2], Pradyuman Singh Rathore[1]

[1]Space Applications Centre, Indian Space Research Organization (ISRO), Ahmedabad-380015, India

[2]Indian Institute of Space Science and Technology, Trivandrum, India

[1] manoj.qit@gmail.com

## Abstract

The AErosol RObotic NETwork (AERONET), established in 1993 with limited global sites, has grown to over 900 locations, providing three decades of continuous aerosol data. While earlier studies based on shorter time periods (10-12 years) and fewer sites (approximately 250) made significant contributions to aerosol research, the vast AERONET dataset (1993-2023) calls for a comprehensive reevaluation to refine global aerosol models and improve satellite retrievals. This is particularly important in light of major environmental changes such as industrialization, land use shifts, and natural events like wildfires and dust storms. In this study, a set of fine and coarse aerosol models called AERONET-Extended (AEROEX) models are developed based on cluster analysis of 30-years AERONET data, analyzing over 202,000 samples using Gaussian Mixture Models to classify aerosol types by season and region. Aerosols are categorized into spherical, spheroidal, and mixed types using particle linear depolarization ratio and fine mode fraction. Four fine-mode aerosol models were derived based on differences in scattering and absorption properties, revealing regional/seasonal variations, particularly in North America, Europe and Asia. Additionally, two coarse-mode aerosol models were identified, separated by their absorbing properties in dust-prone and polluted regions. We performed simulation analysis showing that the new models significantly improve satellite-based aerosol optical depth retrievals compared to widely used dark target aerosol models. The newer aerosol models exhibit distinct optical and physical properties, offering enhanced accuracy in satellite retrievals. A global aerosol model map, generated at 1x1 degree resolution for each season using Random Forest and expert refinement, provides valuable insights for climate and atmospheric studies, improving satellite-based aerosol retrievals at global scale.

# 1. Introduction

Aerosols are tiny liquid, solid, or mixed particles suspended within the Earth's atmosphere, having significantly varying chemical composition and size distribution properties. The aerosols can be of both natural and anthropogenic origin, such as mineral dust from the arid and semi-arid regions, sea salt particles from the oceans, black carbon (BC) from wildfires, bio-aerosols of plant origin, volcanic aerosols, and anthropogenic particle emissions from combustion. Most of the aerosol sources are located near the Earth's surface; therefore, aerosols are mainly concentrated in the lower tropospheric layer. However, particles from desert storms and volcanic eruptions can reach high altitudes and thus can be transported to far distances (Velasco-Merino et al., 2018; Vernier et al., 2018; Wu et al., 2018).

The Intergovernmental Panel on Climate Change (IPCC) in 2021 has classified atmospheric aerosols as short-lived climate forcers (SLCFs) due to their long-term climate impacts (IPCC,2021). Aerosols can affect Earth's radiation budget by scattering and absorbing the solar and longwave radiation, leading to cooling and warming of the Earth's system. . Aerosol acts as cloud condensation nuclei and can warm or cool the atmosphere; therefore, indirectly, they influence the formation of clouds and precipitation by altering cloud properties, such as the number and size of droplets (Twomey, 1977; Albrecht, 1989). They also have a semi-direct effect where the cloud droplets are evaporated due to the absorption of solar radiation by aerosols. Aerosols pose serious health hazards to human beings. Fine aerosols with a diameter of less than 2.5 um, called particulate matter-2.5 (PM2.5), may lead to severe respiratory diseases (Lu et al., 2015; Mukherjee & Agrawal, 2017; Zhang et al., 2014a, 2014b; Kondragunta et al., 2023). They may seriously degrade the atmospheric visibility, affecting air traffic and satellite remote sensing of the earth's surface and resources. Beyond their climatic and health impacts, aerosols play a critical role in atmospheric chemistry, influencing the formation of ozone and other secondary pollutants (Seinfeld & Pandis, 2006). By transporting nutrients and pollutants on land and ocean surfaces, they influence biogeochemical cycles, affecting ecosystems (Mahowald et al., 2005). The socioeconomic consequences of aerosol pollution include significant economic costs associated with healthcare, reduced labor productivity, and damage to infrastructure and agriculture (OECD, 2016). For these reasons, monitoring atmospheric aerosols at global, regional, and local scales is an essential and urgent necessity today. However, due to significant geographical heterogeneity and short atmospheric lifespans (ranging from a few hours to many months), atmospheric aerosol monitoring is an inherently challenging task. The synergistic application of remote sensing techniques employed both from space-based platforms (such as MODIS, OCM-3, etc.) and ground-based networks equipped with sun-photometers AErosol RObotic NETwork (AERONET) has significantly advanced the understanding of geographical aerosol distribution . Satellite-based remote sensing offers a non-intrusive method for acquiring global coverage of aerosol presence (Remer et al., 2005; Remer et al, 2013; Mishra et al., 2018; Mishra et al., 2023, Mishra et al, 2020, Mishra et al, 2021). However, satellite-based aerosol retrievals are susceptible to uncertainties due to the errors in the used surface reflectance and aerosol optical properties for inversion. Ground-based networks such as AERONET, while not capable of providing global coverage, are capable of making precise, detailed, and continuous measurements of aerosol optical properties at key locations (Dubovik et. al., 2002a, 200b; Holben et al., 1998).

Aerosol retrieval algorithms compare the satellite-measured top-of-atmosphere reflectance with the simulated reflectance from radiative transfer models to estimate aerosol parameters like aerosol optical depth (AOD). Having prior knowledge of aerosol models is essential for the radiative transfer calculations in retrieval algorithms and inaccuracies in the assumptions of models and vertical distribution lead to significant errors in the retrieved parameters (Li et al., 2020). Ground observations are capable of providing required information on aerosol optical properties that can be fed into the retrieval processes thereby decreasing uncertainty in satellite-based aerosol retrievals. Most current satellite retrievals provide aerosol optical depth as a primary product that has to be combined with the optical properties of aerosols to address various geophysical problems. Due to the spatio-temporal variability of aerosols, it is not appropriate to assign a set of mean aerosol physical and chemical properties to a given location based on a long-term average. However, analyzing a large database of ground-based aerosol measurements collected over short timescales (a few hours) across many years can help determine the likelihood of identifying specific aerosol types in particular seasons and regions (Omar et al., 2005; Levy et al., 2007). Several satellite-based aerosol retrieval algorithms have utilized global and regional aerosol models derived from the long-term average of AERONET measurements. Aerosol retrieval algorithms like Dark-target, Deep-blue, BAER, etc., necessitate selecting the aerosol model that best represents the aerosol conditions in a specific region seasonally. It is to be noted that, MODIS Dark Target retrieval algorithms combine a unique fine-dominated aerosol model (selected from three predefined fine models) with a coarse-dominated one for each retrieval process. Therefore, the classification of aerosols based on the analysis of large datasets of ground-based aerosol measurements is crucial. This approach ensures that the retrieval algorithms can more accurately account for the regional and seasonal variability of aerosols, thereby improving the accuracy of satellite-based aerosol retrievals and enhancing our ability to address various geophysical and climate-related problems. Additionally, such classification will also provide valuable inputs for addressing several geophysical or climate change studies that require both AOD and aerosol properties, given that most satellites provide accurate AOD measurements only.

Several studies have explored the use of optical and microphysical properties of aerosols to distinguish distinct aerosol types. Ground-based aerosol networks like AERONET (Holben et al., 1998) provide long-term data on aerosol optical and micro-physical properties, which includes the spectral aerosol optical depth (AOD), complex refractive index, particle size distribution, single-scattering albedo (SSA), asymmetry parameter (ASYM), etc. The comprehensive knowledge of these properties of the aerosol types is essential for elucidating the mechanisms of aerosol radiative forcing and improving the accuracy of satellite retrieval algorithms.

In the past, the commonly utilized approach for identifying different aerosol classifications relies on applying specific threshold criteria to various microphysical and optical properties of aerosols. Additionally, source apportionment techniques also aid in classifying aerosols based on their emission sources. The studies, listed in Table 1, have employed empirical threshold methods for aerosol classification. However, the limitations of threshold methods lie in their inability to classify aerosols with complex chemical compositions accurately. Most of these works are based on older data collected before 2010 over a short period of the order of 10 years or less. Some regional studies such as Shin et al. (2019) have used relatively newer data collected in China for 20 years from 1993-2017.

**Table 1.** Studies on Aerosol Type Classification using Empirical Threshold Methods. Here, SSA, ASYM, FMF, AAE, AE and PLDR stands for single scattering albedo, asymmetry parameter, fine mode fraction, angstrom absorption exponent, angstrom exponent, and particle linear depolarization ratio, respectively.

| Study Region & Period | Parameters (Microphysical/Optical Aerosol Properties) | Aerosol Types | Reference |
|---|---|---|---|
| Worldwide Locations 1993 –1999 | Wavelength Dependence of AE | i. Biomass Burning<br>ii. Urban/Industrial<br>iii. Desert Dust | Eck et al., 1999 |
| Worldwide Locations 1993 – 2000 | SSA, Particle size Distribution and Index of Refraction | i. Biomass Burning<br>ii. Urban/Industrial<br>iii. Desert Dust<br>iv. Marine Aerosols | Dubovik et. al., 2002 |
| East Asia | SSA and FMF | i. Dust<br>ii. Mixture<br>iii. Non-absorbing<br>iv. Black carbon | Lee et al., 2010 |
| Worldwide Locations 1993 – 2000 | AAE | i. Urban/Industrial<br>ii. Biomass Burning<br>iii. Sahara Dust | Russel et al., 2010 |
| Worldwide Locations 1999 – 2010 | SSA, AAE, AE and FMF | i. Dust<br>ii. Mixed<br>iii. Urban/Industrial<br>iv. Biomass Burning | Giles et al., 2012 |
| East Asia 2003 – 2013 | SSA Spectral Curvature | i. Black Carbon<br>ii. Dust<br>iii. Mixed | Li et al., 2015 |
| China 2010 – 2017 | SSA, FMF and AAE | Eight Aerosol Types of Fine, Mix and Coarse types. | Che et al., 2019 |
| East Asia 1993 – 2017 | PLDR and SSA | i. Pollution-dominated Mixture<br>ii. Dust-Dominated Mixture<br>iii. Pure Dust<br>iv. Non-absorbing<br>v. Weakly absorbing<br>vi. Moderately absorbing<br>vii. Strongly Absorbing | S.-K. Shin et. al., 2019 |
| Europe, Middle East, North Africa, and Arabian Peninsula 2008 – 2017 | SSA, FMF and AE | Ten Aerosol Types of Fine, Mix, Coarse and Other. | Logothetis et al., 2020 |

In contrast to threshold-dependent approach, clustering analysis serves as an unsupervised machine learning approach utilized for classifying unlabeled and uncategorized datasets based on specific measured parameters into distinct classes with similar characteristics. The objective of clustering is to minimize within-cluster variability and maximize between-cluster differences thereby ensuring that the data points within each cluster are as similar as possible while the clusters themselves are as distinct from each other as possible. Some of the widely used clustering techniques are K-means (Partitioning method), Balanced Iterative Reducing and Clustering using Hierarchies (BIRCH), Density-Based Spatial Clustering of Applications with Noise (DBSCAN), etc. Several studies, as outlined in Table 2, have applied clustering analysis, leveraging aerosol properties as features for classification. Among these three studies are regional (China and Australia) that essentially cover only locally sourced aerosols and thus do not represent the true picture of aerosol models at the global scale. None, of these studies has considered the sphericity of aerosol particles while doing cluster analysis. Only two studies by Levy et al. (2007) and Omar (2005) were performed at global scale.

**Table 2.** Studies of Aerosol Type Classification using Clustering Analysis. Here, SSA and ASYM stands for single scattering albedo and asymmetry parameter, respectively.

| Study Region & Period | Parameters (Microphysical/Optical Aerosol Properties) | Aerosol Types | Reference |
|---|---|---|---|
| Worldwide Locations 1993 – 2002 | 26 Parameters of size distribution, chemical composition, and optical properties | i. Desert Dust<br>ii. Biomass Burning<br>iii. Urban Industrial Pollution<br>iv. Rural Background<br>v. Polluted Marine<br>iv. Dirty Pollution | Omar et al., 2005 |
| Worldwide Locations 1993 – 2005 | SSA and ASYM | i. Dust<br>ii. Non – Absorbing<br>iii. Moderately – Absorbing<br>iv. Absorbing | Levy et. al., 2007 |
| Australia | 12 Optical properties | i. Aged smoke<br>ii. Fresh smoke<br>iii. Coarse dust<br>iv. Super-absorptive aerosol | Qin and Mitchell, 2009 |
| China 1998-2017 | 25 aerosol microphysical properties | i. Desert dust<br>ii. Scattering mixed<br>iii. Absorbing mixed<br>iv. Scattering fine | Zhang and Li, 2019 |
| China 2001-2019 | 22 Optical properties | i. Carbonaceous-black carbon<br>ii. Carbonaceous-brown Carbon<br>iii. Dust-Desert<br>iv. Dust-City<br>v. High humidity & high pollution<br>vi. High-humidity & low-pollution<br>vii. Low-humidity & high-pollution<br>viii. Low-humidity & low-pollution | Fan et. al., 2021 |

The aerosol models presently utilized in satellite algorithms like MODIS were established in 2007. Levy et al. (2007) developed these models through subjective cluster analysis using aerosol optical properties—specifically, Aerosol Optical Depth (AOD) at 550nm, Single Scattering Albedo (SSA) at 675nm, and asymmetry parameter (ASYM) at 440nm—derived from AERONET data retrievals. They identified the dominant aerosol type at each site and extrapolated it to create seasonal 1° × 1° global maps of expected aerosol types. Specifically, three fine-mode aerosols and one non-spherical coarse-mode aerosol model were developed. Omar et al. (2005) has classified the AERONET measurement into 6 models without considering fine and coarse dominance and sphericity as classification criteria. Since satellite data provides only a few spectral bands (2 or 3) for inversion, therefore, Levy's (2007) models were more appropriate as they allow the external mixing of the fine and coarse model identified for the region leading to the use of temporally and spatially dynamic aerosol optical properties for satellite-based aerosol retrievals. Since then, the AERONET network has expanded, and in recent years, significant changes in the microphysical and optical properties of aerosols are expected due to increased anthropogenic activities and land use changes. Additionally, at that time, AERONET data did not provide any information on the sphericity of particles.

Recently, with the increased utilization of active aerosol remote sensing using LIDAR, researchers have been able to obtain detailed and quantitative information crucial for aerosol classification. This approach particularly leverages a shape-sensitive parameter known as the particle linear depolarization ratio (PLDR or $\delta p$) (Shin et al., 2019). Several studies have successfully identified aerosol types by integrating LIDAR-derived PLDR with aerosol optical and microphysical properties inferred from AERONET observations. However, the integration of polarization LIDAR with AERONET observations is limited by the scarcity of AERONET stations equipped with additional instrumentation, such as well-calibrated polarization-sensitive micro-pulse LIDAR or aerosol polarization LIDAR, capable of providing PLDR measurements. In Version 3 of the AERONET retrieval methodology, spectral PLDR is provided as a standard inversion product, offering valuable insights for a more comprehensive understanding of aerosol characteristics, shapes, and enhancing the accuracy of aerosol classification.

The present study advances the methodology introduced by Levy et al. (2005) for developing new aerosol models using enhanced microphysical and optical properties sourced from the AERONET database. This study integrates data on particle shape (Spectral PLDR) and employs machine-learning techniques, in addition to parameters previously utilized for clustering analysis. These methods aim to identify distinct aerosol models based on location and season. In addition, this study utilizes aerosol data spanning three decades, significantly surpassing the dataset used in any of the past studies incorporating data from more than 900 sites. This extensive dataset enables the inclusion of recent changes (specifically in the last two decades) in aerosol spatio-temporal distribution. Subsequently, the spatial-temporal distribution of the derived global aerosol models is identified through a supervised classification method and represented using 1° × 1° global maps. This approach facilitates a better understanding of the spatiotemporal variability of global aerosol characteristics and their underlying drivers. The implication of using new aerosol models on aerosol retrieval is also presented based on simulation study.

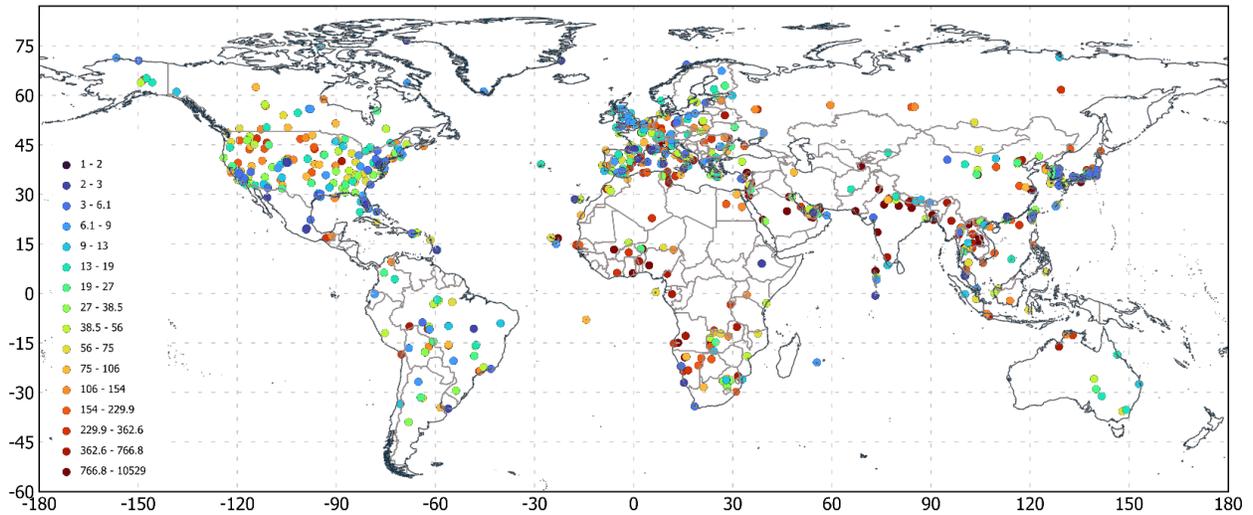

Figure 1. The location of 989 AERONET station for which inversion products are available and are used in present cluster analysis. The color of the dots represents number of samples available at each site.

## 2. Dataset used

### 2.1 AERONET data.

The AERONET (AErosol RObotic NETwork) is a global network of sun photometers that provides long-term continuous data on the microphysical, optical, and radiative properties of aerosols, obtained from direct sun measurements and inversion of spectral sky radiances. The direct Sun measurements are available at a temporal resolution of 15 min or better during the daytime at eight spectral bands between 340nm to 1020nm (standard wavelengths: 440 nm, 670 nm, 870 nm, 940 nm, and 1020 nm). Almucantor and Sky scan measurements, at 440 nm, 670 nm, 870 nm, and 1020 nm, are the two sky scan scenarios available in the AERONET sun photometer, providing measurements once every hour. In Almucantor sky scans, radiances are measured at azimuthal angles with respect to the sun, and in principle plane sky scans, they are measured at scattering angles in the solar principle plane away from the Sun.

The AERONET retrieves the aerosol microphysical properties including particle size distribution (Volume concentration, volume median radius, standard deviation, and effective radius for total, fine, and coarse aerosol modes in 22 size bins ranging from 0.05 to 15µm) as well as the complex refractive indices. Additionally, the inversion algorithm calculates the aerosol optical and radiative properties using these retrieved microphysical properties at sky radiance wavelengths. This includes parameters such as single scattering albedo, phase function for 83 scattering angles, asymmetry parameter, and spectral fluxes.

The AERONET measurements from both sun and sky observations undergo rigorous calibration and cloud screening procedures. Version 3 AOD data are available in three quality levels: Level 1.0 (unscreened), Level 1.5 (cloud-screened and quality-controlled), and Level 2.0 (quality-assured). Identification of aerosol types in the present study utilizes the AERONET level 2.0 version 3 inversion products (referred to as "L2_v3A" products in this paper). These products are derived from the climatology of almucantar retrievals from all global AERONET Sun photometer sites processed as of July 2023. Figure 1 shows the location of 989 AERONET stations for which data is downloaded and used in the present work. The color of the dots in Figure 1 represents the number of samples available at each AERONET site.

## 2.2 Parameters used for cluster analysis

Aerosol particles can be classified based on their distinct tendencies to absorb and scatter incident solar radiation, the angular distribution of scattered light, and their varying shape and size. These properties are crucial for classification due to their significant impact on the radiative transfer of solar energy from the top-of-atmosphere to the Earth's surface and back to space. Understanding this solar energy transfer is essential for addressing several geophysical problems, including climate change studies, and for advancing satellite-based remote sensing techniques. The parameters utilized for the clustering analysis include aerosols' microphysical and optical properties. Specifically, five inversion parameters namely: PLDR, SSA, ASYM, and volume concentration of fine and coarse modes, are used for classification.

### 2.2.1 Volumetric Fine Mode Concentration

Instead of the direct use of volume concentration of fine and coarse modes for cluster analysis, the volumetric fine mode fraction (VFMF) is used in the present cluster analysis. To calculate VFMF following relation is used:

$$VFMF = \frac{C_v^f}{C_v^f + C_v^c}, \tag{1}$$

where $C_v^f$ and $C_v^c$, respectively, are the volume concentration of fine and coarse modes.

### 2.2.2 Particle Linear Depolarization Ratio

Unlike version 2 of the AERONET inversion product, version 3 data also includes spectral Particle Linear Depolarization Ratio (PLDR). PLDR can provide information about the shape and composition of atmospheric aerosols. It is defined as the ratio of the intensity of the backscattered light in the perpendicular polarization state ($P_\perp$) to that in the parallel polarization state ($P_\parallel$), given by:

$$PLDR = P_\perp / P_\parallel, \tag{2}$$

In the case of LIDAR-based sensing, a linearly polarized laser beam is sent into the atmosphere and the polarization-sensitive measurement of the reflected signal is performed. The reflected signal is linearly polarized if interacting particles are spherical. However, for non-spherical particles, the reflected signal has a different plane of polarization having a non-zero perpendicular polarization component. High Spectral resolution LIDAR measures PLDR directly using specialized depolarization channels. In Version 3 of the AERONET retrieval, Spectral PLDR at wavelength ($\lambda$) 440, 670, 870, and 1020 nm are provided as standard inversion products. The calculation involves the estimation of diagonal elements $F_{(11,\lambda)}(r, n, \theta)$ and $F_{(22,\lambda)}(r, n, \theta)$ at the scattering angle $\theta$=180° of the Muller scattering matrix (Bohren and Huffman, 1983) and is calculated as:

$$PLDR = \frac{1 - F_{(11,\lambda)}(r,n,180°) / F_{(11,\lambda)}(r,n,180°)}{1 + F_{(11,\lambda)}(r,n,180°) / F_{(11,\lambda)}(r,n,180°)} \tag{3}$$

The Muller matrix elements $F_{(11,\lambda)}(r,n,\theta)$ and $F_{(22,\lambda)}(r,n,\theta)$ strongly depend on the angular and spectral distribution of the radiative intensity, measured by AERONET. They are computed using the retrieved complex refractive indices and particle size distributions. $F_{(11,\lambda)}(r,n,\theta)$ is proportional to the flux of the scattered light for un-polarized incident light. Another crucial input parameter for the PLDR retrieval is the aspect ratio distribution, which signifies the ratio between a particle's longest and shortest axes; this distribution is maintained as a fixed parameter in the AERONET model due to the nearly equivalent scattering characteristics of spheroid mixtures (Lee et al., 2010).

PLDR value ranges between 0.30 and 0.35 indicating non-spherical (spheroids) coarser-sized particles like mineral dust or volcanic ash, while near-zero values indicate the presence of fine-sized spherical particles. Intermediate values suggest a mixture of spherical and non-spherical particles. The analysis of PLDR measurements can help estimate the contribution of these two distinct types of particles (Shin et al., 2019). Noh et al. (2017) reported a strong correlation between PLDR values derived from AERONET inversion products and those obtained from LIDAR measurements. Several studies conducted in this regard indicate that the PLDR values retrieved from AERONET at wavelengths 870 and 1020 nm are the more reliable (Shin et al., 2019). Therefore, in the present analysis, we have opted to utilize the PLDR at 870 nm to take the size and shape of the aerosol particles into account while classification.

### 2.2.3 Single Scattering Albedo (SSA)

Single scattering albedo (SSA) is an aerosol optical property calculated as the ratio of scattering efficiency ($Q_{sca}$) to total extinction efficiency ($Q_{ext}$) of the particle:

$$SSA\ (\lambda) = \frac{Q_{sca}((\lambda))}{Q_{ext}((\lambda))} \tag{4}$$

SSA depends on the scattering and absorption strength of the particles. Perfectly scattering aerosols have SSA very close to unity whereas lower SSA values indicate the presence of absorbing particles. The significance of SSA is evident in its role in estimating direct radiative forcing and characterizing aerosol types based on their optical properties. Analyzing SSA alongside other microphysical and optical properties allows aerosols to be categorized into distinct groups. In the present study, AERONET-retrieved SSA values at 675 nm are utilized for clustering analysis.

### 2.2.4 Asymmetry Parameter (ASYM)

The angular distribution of light scattered by a particle is represented by the asymmetry parameter (ASYM). It is calculated as the average of the cosine of the scattering angles weighted by the phase scattering function, $P_a(\lambda, \theta)$, given by equation:

$$ASYM = \frac{\int_0^\pi \cos\theta\ P_a(\lambda,\theta)\ d(\cos\theta)}{\int_0^\pi P_a(\lambda,\theta)\ d(\cos\theta)} \tag{5}$$

Theoretically, the value of ASYM can range from -1 to +1. Aerosols with larger particle sizes tend to have higher asymmetry parameters. The asymmetry parameter plays an important role in determining aerosol direct radiative forcing and radiation pressure (K. Ehlers and H. Moosmuller, 2023). In the present study, AERONET-retrieved ASYM values at 440 nm are utilized for clustering analysis.

**2.3 Other data sets**

*MERRA-2 model reanalysis data.*

MERRA-2 (Modern-Era Retrospective analysis for Research and Applications, Version 2) model provides various atmospheric and land surface variables, including dust density and wind data. Here MERRA-2 tavgM_2d_aer_Nx:2d, Monthly-mean, Time-averaged, Single-Level, Assimilation, Aerosol Diagnostics V5.12.4 (M2TMNXAER) collection is used for the analysis of dust density at global scale. This collection offers assimilated aerosol diagnostics such as column mass density of different aerosol components (black carbon, dust, sea salt, sulfate, and organic carbon), surface mass concentrations and total optical thickness at 550nm. The monthly averaged dust column mass density product are sourced from MERRA-2 M2TMNXAER collection, available through NASA's Giovanni web portal that provides detailed information on dust mass density integrated over the entire atmospheric column. This data is available at spatial resolution of 0.5 x 0.625 degrees. For period from January 2017 to December 2023, monthly product data were utilized to compute seasonal averages. This process involved aggregating the monthly data derive seasonal means, facilitating an analysis of temporal variations and seasonal patterns in dust distribution and origination across the globe. MERRA-2 tavgM_2d_flx_Nx:2d, Monthly-mean, Time-averaged, Single-Level, Assimilation, Surface Flux Diagnostics V5.12.4 (M2TMNXFLX) collection is used to analyze the seasonal wind patterns across globe. This collection offers several assimilated surface flux diagnostics at spatial resolution of 0.5 x 0.625 degrees. The surface wind speed eastward and northward components are downloaded from NASA's Giovanni web portal and seasonal averages of wind speed and direction are computed using data from January 2017 to December 2023. The combined use of seasonal averages of dust columnar mass density and wind patterns offers valuable insights into the origins and transport of dust aerosols. This approach helps identify areas affected by the dust transport and can be correlated with observed optical and microphysical properties of aerosols, as revealed through the cluster analysis of ground observations by AERONET stations across globe.

*MODIS Active Fire product*

In addition to the dust column mass density and wind pattern data, seasonal averages of cumulative fire counts are generated to complement the analysis. The Near real-time Moderate Resolution Imaging Spectroradiometer (MODIS) Thermal Anomalies / Fire locations - Collection 61 processed by NASA's Land, Atmosphere Near real-time Capability for EO (LANCE) Fire Information for Resource Management System (FIRMS), using swath products (MOD14/MYD14) is used in present work to discuss the distribution and origin of fine mode carbonaceous aerosols. This product is freely downloadable from the Fire Information for Resource Management System (FIRMS) website https://firms.modaps.eosdis.nasa.gov. The thermal anomalies / active fire represent the center of a 1 km pixel flagged by the MODIS MOD14/MYD14 Fire and Thermal Anomalies algorithm (Giglio et al., 2003) as containing one or more fires within the pixel. The data also includes information on the confidence level of fire detection. Here detections with very good confidence better than 75% are only used and the rest are rejected for the analysis. The vector data is processed and rasterized at 10km spatial resolution with each pixel containing a spatially and temporally cumulative number of fire counts, thereby giving seasonal raster images of total fire detection over the Indian landmass. For temporal aggregation, data from January 2017 to December 2023 is used. This data provides insights into major sources of carbonaceous aerosols, which are significant contributors to atmospheric fine-mode particulate matter across the globe. This combined data facilitates a comprehensive discussion and understanding of aerosol sources and their global distribution, enhancing the results of cluster analysis of ground-based AERONET observations across the globe.

## 3. Method

### 3.1. Gaussian Mixture Model (GMM) Clustering

The studies tabulated in Table 2 used the K-means method for the analysis of ground-based measurement of aerosol optical properties. K-mean is a popular unsupervised ML algorithm used for partitioning a dataset into K clusters. It works well for many types of data, but may not be the best method for all cases.

Gaussian Mixture Models (GMM) clustering which is an unsupervised ML algorithm shows several advantages over K-means clustering especially in cases where modeling of complex data distribution is involved. GMMs are probabilistic machine-learning (ML) techniques for unsupervised clustering and density estimation that extend beyond the simplistic assumptions made by K-means. In GMM, clusters are formed by representing the dataset's probability density function (PDF) as a superposition of multiple Gaussian distributions, each characterized by its own mean and covariance matrix. Mathematically, each cluster k in a GMM is modeled as a multivariate Gaussian distribution:

$$\mathcal{N}(x|\mu_k, \sigma_k) = \left[(2\pi)^{d/2}|\sigma_k|^{1/2}\right]^{-1} exp\left(-\frac{1}{2}(x-\mu_k)^T \sigma_k^{-1}(x-\mu_k)\right), \quad (6)$$

where $\mu_k$ is the mean vector and $\sigma_k$ is the covariance matrix of the k-th Gaussian component. The overall probability density function for the GMM is a weighted sum of these Gaussian component densities:

$$p(x) = \sum_{k=1}^{K} \pi_k \mathcal{N}(x|\mu_k, \sigma_k), \quad (7)$$

where $\pi_k$ are the mixture weights that sum to unity. The parameters of GMM are estimated using expectation-maximization (EM) algorithms, which iteratively calculate the posterior probabilities and update the parameters until convergence is achieved.

Unlike K-Means, GMM can model clusters of various shapes by using different covariance structures. In contrast to hard assignment in K-means, GMM permits soft clustering, wherein data points are assigned to multiple clusters with varying degrees of probabilities. This probabilistic framework is particularly useful for handling overlapping clusters and complex data distributions.

For clustering analysis of aerosol optical and physical properties measured globally at more than 900 sites over a long time domain of 3 decades, the advantages of GMM are even more pronounced. Aerosol data exhibits heterogeneity due to varying aerosol sources, environmental conditions, and atmospheric processes. For example, an urban environment can have complex mixtures of aerosol types from traffic, industrial activities, and natural sources. The GMM's ability to model these diverse distributions makes it more suitable for accurately reflecting the complexity of aerosol measurements. The probabilistic nature of GMM can identify the transition between different aerosol types, providing deeper insight into aerosol dynamics and interactions. Given these advantages, in the present study, the use of GMM is preferred over K-means to have interpretable cluster outcomes.

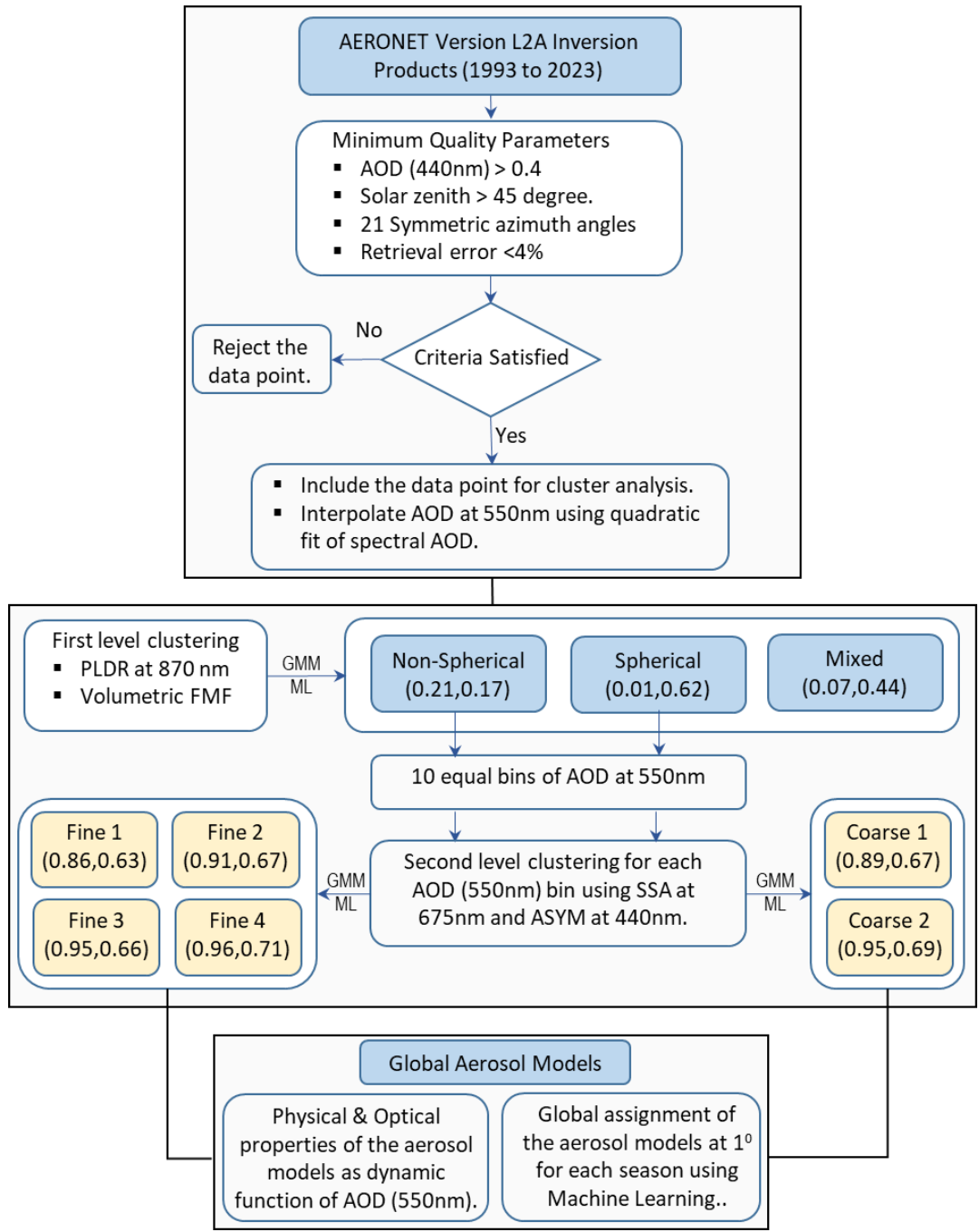

Figure 2. Flowchart of the clustering analysis. Here, AOD, PLDR, FMF, SSA, ASYM, GMM and ML refer to Aerosol Optical Depth, Particle Linear Depolarization ratio, Fine Mode Fraction, Single Scattering Albedo, Asymmetry Parameter, Gaussian Mixture Model and Machine Learning, respectively. The values in the bracket indicate the average PLDR and Volumetric FMF for the non-spherical, spherical and mixed classes obtained from first level cluster analysis. The values in the bracket for the global aerosol models obtained after the second level clustering analysis indicate the average SSA and ASYM.

## 3.2 Clustering analysis with AERONET data

In the present work, we have used GMM for the clustering analysis of aerosol data measured by AERONET over 3 decades globally. The clustering analysis utilizes the retrieved aerosol properties from the "L2_v3A" products. The AERONET retrievals are filtered using the minimum quality parameters outlined in the Quality Manual of AERONET to ensure the highest accuracy of the retrieved aerosol information. The quality criteria for selecting the valid data are AOD (440nm) greater than 0.4, solar zenith angle greater than 45°, 21 symmetric left/right azimuth angles, and radiance retrieval error less than 4%. The resulting dataset comprised 202735 L2_v3A Almucantar Retrievals.

Figure 2 illustrates the flowchart detailing the Aerosol Clustering Analysis methodology adopted in this study. Here a two-level classification scheme is adopted. At the first level, we classified the entire dataset into three classes namely: spherical fine-dominated, non-spherical or spheroid coarse-dominated, and mixed aerosol samples. To do this PLDR and VFMF are used in the GMM ML algorithm. The decision to attempt 3 clusters at the first level is subjectively based on the distribution of PLDR. Figure 3 illustrates the histogram of PLDR, indicating three distinct aerosol types in terms of shapes that possibly represent the dataset's spherical, mixed, and non-spherical (spheroids) classes. Additionally, the density plot of PLDR at 870nm and VFMF (Figure 4) identifies two dense regions with "low PLDR and high VFMF" and "high PLDR and low VFMF" emphasizing the spherical fine-sized and non-spherical coarse-sized categories, respectively. Thus, the first clustering was carried out using the GMM unsupervised technique to differentiate the three classes, utilizing PLDR at 870nm and VFMF as the features. The resulting three clusters comprise 91301 spherical, 54508 non-spherical, and 56926 mixed retrievals. Assuming that the physical and optical properties of the mixed particles can be obtained by external mixing of the properties of the spherical and non-spherical particles, we proceeded with the second-level clustering separately for spherical and non-spherical groups, utilizing the SSA as an absorbing strength parameter and the asymmetry parameter describing the angular scattering of light.

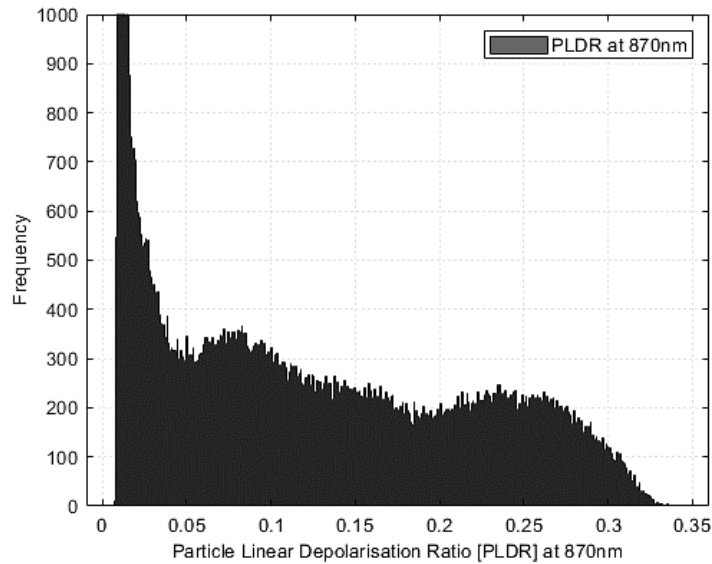

Figure 3. Histogram of Particle Depolarization Ratio (PLDR) at 870nm. (Note the limit of the Y axis is 20000, it is limited in the Figure to visualize the three modes distinctly).

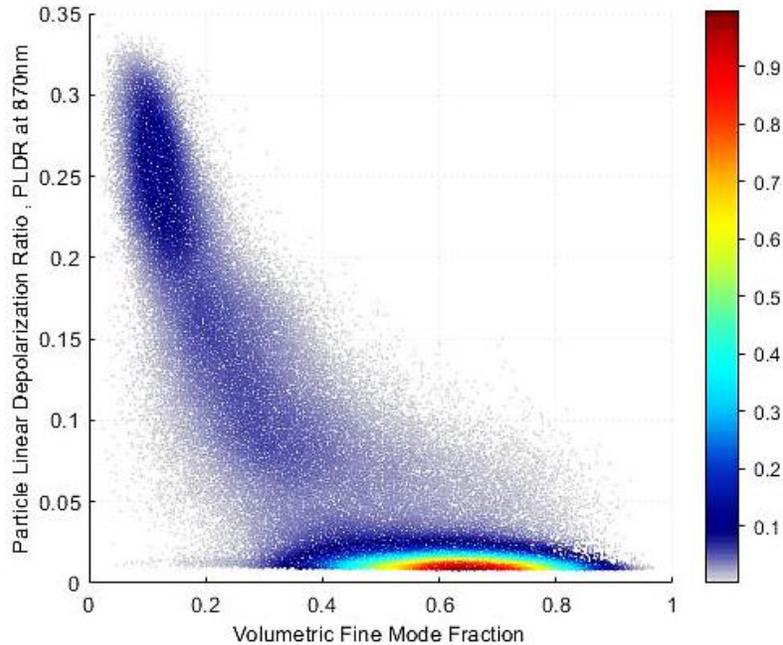

Figure 4. Density Plot of Volumetric Fine Mode Fraction and Particle Depolarization Ratio (PLDR) at 870nm.

The optimum number of clusters in the spherical and non-spherical categories is decided using the silhouette score. The silhouette coefficient tells how well a data point belongs to its assigned cluster. The overall silhouette score is obtained by averaging the coefficients for all the data points. A high average silhouette coefficient indicates good clustering. Figure 5 (left panel) shows the Silhouette score for the Spherical datasets. The peaks are observed at clusters 2, 4, and 9. While results with 2 clusters will be insufficient to identify the distinct types of spherical classes, 9 will be too large. Thus, an optimum number of clusters for the spherical categories is taken as four. For non-spherical datasets (Figure 4, right panel), the Silhouette score peaks at cluster numbers 2 and 7. Since seven clusters are inappropriate, the non-spherical retrievals are to be classified into two classes.

To account for variations in aerosol optical properties with aerosol concentration, AOD at 550nm is calculated by performing a quadratic fit on the spectral AOD of the AERONET retrievals for the spherical and non-spherical particles in log space. Representing aerosol properties as a function of optical thickness leads to dynamic aerosol models, which help parameterize aerosols in climate models and satellite remote sensing. This relationship provides insight into how aerosol concentration and aerosol processes, including aging, particle size, and chemical transformations, influence the optical behavior of the aerosols (Remer and Kaufman 1998; Remer et al. 1998). The samples are divided into ten equal bins of AOD at 550nm to identify the dynamic variability of aerosol properties with respect to the aerosol optical thickness. A second level of GMM clustering analysis is then conducted for all τ bins of the spherical and non-spherical groups separately. The cluster results from separate AOD bins are combined to describe the dynamical properties of aerosol types collectively. Clustering analysis is performed using SSA at 0.67μm and the asymmetry parameter (ASYM) at 0.44μm. Separate wavelengths are adopted for both optical properties because dust particles exhibit strong absorption of solar radiation in the blue spectral range and using SSA at 440 nm may result in the undesirable creation of dust clusters. Larger-sized aerosols can be effectively differentiated by the asymmetry parameter at shorter wavelengths, i.e., 440 nm.

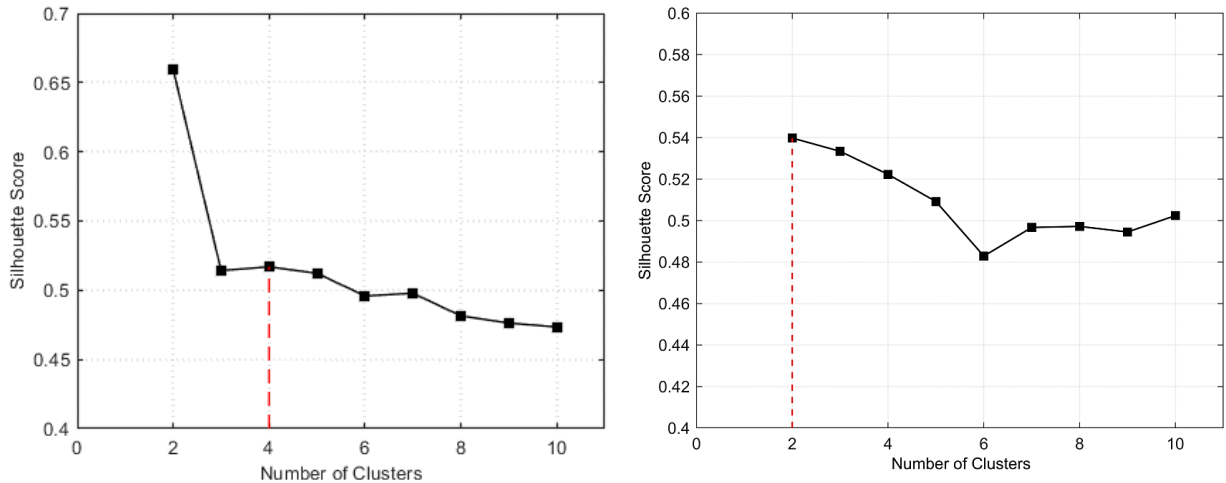

Figure 5. Identification of the Optimum Number of Clusters for spherical retrievals (left), and for non-spherical retrievals (right).

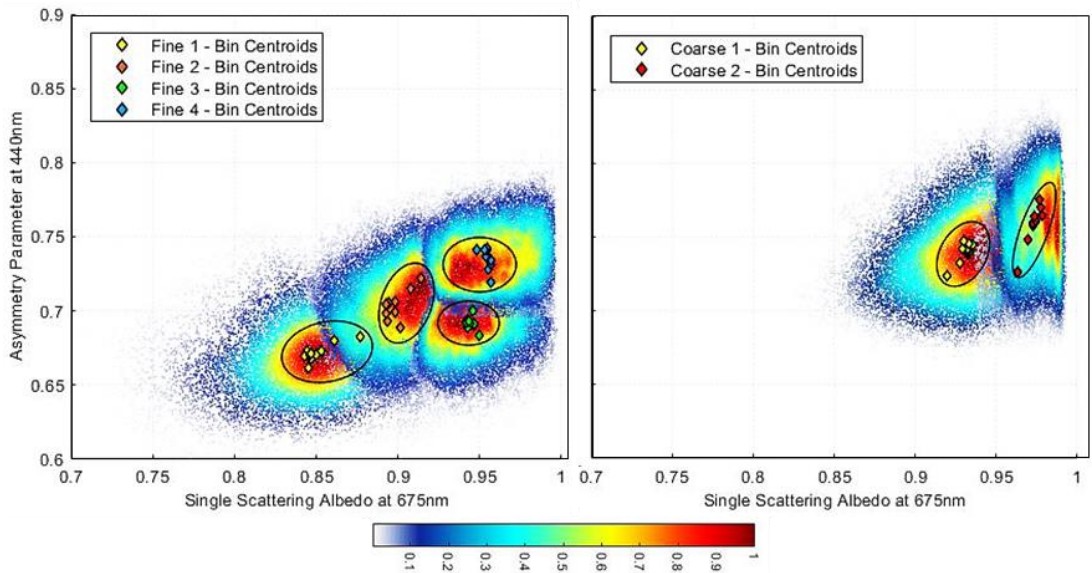

Figure 6. (a) Density Plot of Single Scattering Albedo at 675nm and Asymmetry Parameter at 440nm to differentiate the Spherical Clusters: Fine 1, Fine 2, Fine 3 and Fine 4. The Cluster centroids for 10 aerosol optical depth bins are also shown. (b) Same to (a) but for Spheroid Clusters: Coarse1 and Coarse2.

## 4. Results and Discussion

### 4.1 Output Clusters or Aerosol Models

As a result of cluster analysis, a set of fine- and coarse-mode dominated aerosol models has been identified. Based on the extended spatiotemporal data acquired by AERONET, these models represent the major fine- and coarse-mode aerosol types at a global scale, and we have named them the AERONET-Extended (AEROEX) aerosol models. Figure 6 (a) and (b) depicts the clusters identified for the spherical and non-spherical categories, respectively. The cluster centroids for 10 AOD bins are also shown. Four distinct fine-mode dominated spherical aerosol types are identified as Fine-1, Fine-2, Fine-3, and Fine-4 while Coarse-1 and Coarse-2 are the coarse-mode dominated non-spherical clusters. The density plot shows that the maximum population in each identified cluster (yellow and red shades) are well separated, while the variability of aerosol optical properties (SSA-675nm and ASYM-440nm) with AOD within each cluster is also evident by the distribution of cluster centroid for each AOD bin.

### 4.2 Micro-Physical and Optical Properties of the Aerosol Models

The microphysical properties of the derived models are obtained by averaging the properties within each aerosol cluster. The particle distribution curves for each aerosol model for different bins of AOD at 550nm are illustrated in Figure 7. The particle distribution function shows a strong dependency on the value AOD at 550nm. The positive shift in the position of peaks of distribution function with increasing AOD values is more evident for fine-dominated aerosol models, while for coarse-dominated models no significant shift in peak location is observed. This shows that on average with increased aerosol loading the population of particles with larger radii increases for the case of four fine-mode dominated aerosol models identified here. Volume concentrations i.e., the area under the distribution curves show a strong positive correlation with AOD value both for fine and coarse models. The difference in the width of distribution curves for different models is also apparent. Note the drastic difference in distribution curves for four fine and two coarse mode models. The observed dependency of microphysical properties on AOD at 550nm led us to represent size distribution parameters as a function of AOD at 550nm for each aerosol model. Table 3 quantitatively summarizes the particle size distribution parameters (median radius of the volume size distribution ($r_v$), standard deviation ($\sigma = \ln(\sigma_v)$), and Volume concentration ($V_o$) ) and the complex refractive indices of the derived cluster models along with the "Continental Model" (Lenoble and Brogniez, 1984) as a function of AOD at 550nm.

Table 4 shows the optical properties of the derived aerosol models compared to the Continental Model, which includes the SSA, ASYM, and extinction coefficient ($\alpha$) all calculated for AOD=0.5 at 550nm. The values at 550nm are highlighted by bold fonts. It is evident from Table 4 that Fine-1 and Fine-2 models having SSA=0.86 and 0.93 exhibit strong and moderate absorbing strength, respectively, while Fine-3 and Fine-4 models both having SSA > 0.95 are non-absorbing in nature. Even though Fine-3 and Fine-4 models show similar SSA, they are significantly different in terms of angular scattering capabilities represented by different ASYM of 0.65 and 0.71, respectively. Specifically, Fine-4 is more strongly forward scattering than Fine-3 and the other two fine models. All four fine-mode aerosol models have significantly different extinction efficiencies. Coarse-1 and Coarse-2, the two non-spherical coarse mode models are primarily different in their absorbing strength with SSA values of 0.897 and 0.95, respectively, while having similar values of asymmetry parameters. Coarse-2 primarily represents non-absorbing spheroid aerosols similar to mineral dust, while Coarse-1 model interestingly shows significant absorbing strength even stronger than moderately absorbing Fine-2 model.

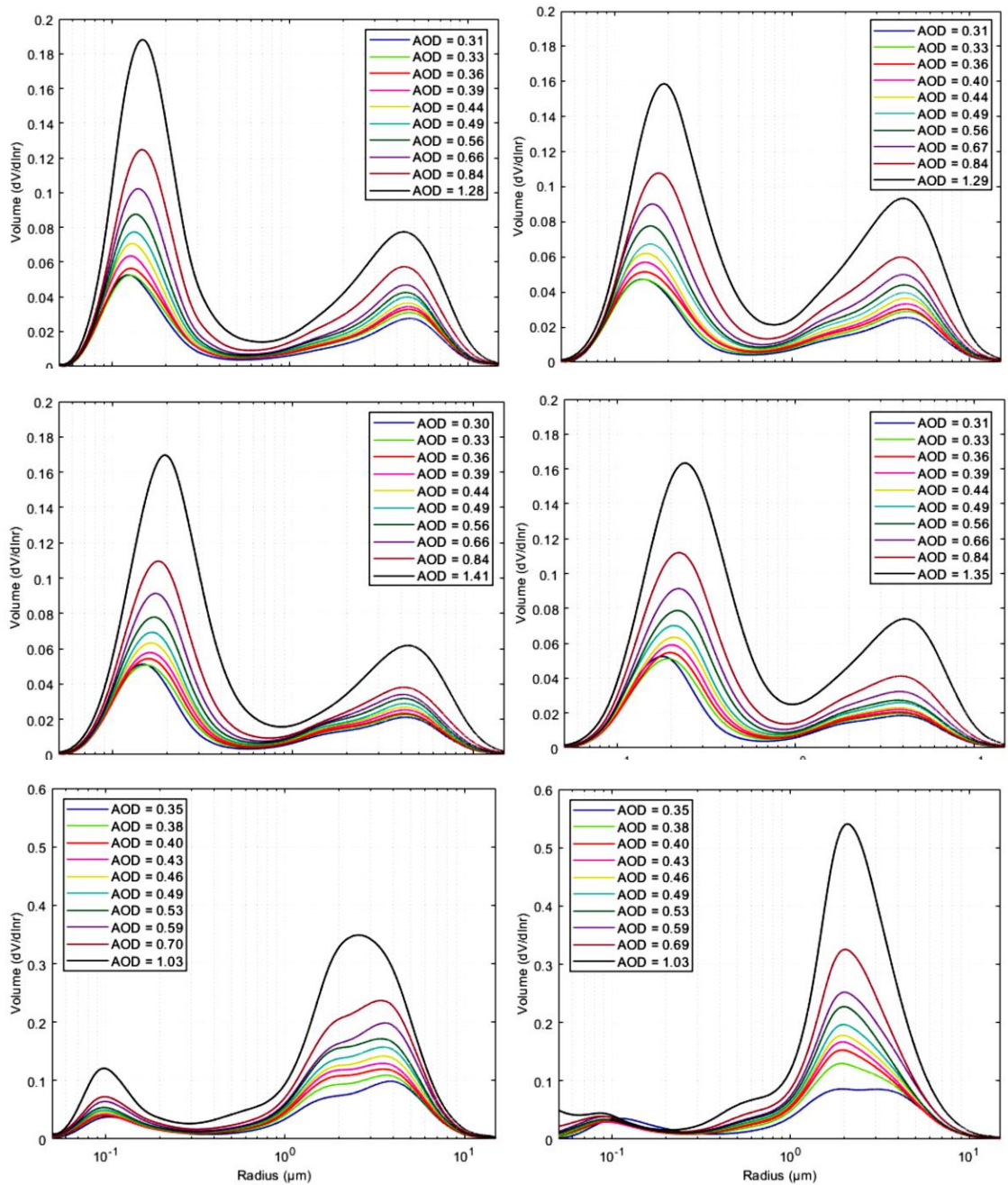

Figure 7. Particle Size Distribution Curve for the Fine Dominated Aerosol Models (Fin-1, Fine-2, Fine-3, and Fine-4) and Coarse Dominated Aerosol Models (Coarse 1, and Coarse 2) as a function of τ.

Table 3. Particle Size Distribution Parameters and Refractive indices of Aerosol Models. Symbols $\tau$, $r_v$, $\sigma$, and $V_0$ refer to aerosol optical depth at 550nm, median radius, standard deviation, and Volume concentration, respectively. The refractive index is represented by real component n and Imaginary component k. The properties for continental model tabulated here are from Lenoble and Brogniez, (1984).

| Model | Mode | $r_v$ (μm) | σ | $V_0$ (μm³/μm²) | Refractive Index (n-ki) for wavelengths λ=0.44,0.55,0.675,0.87,1.02 nm |
|---|---|---|---|---|---|
| Fine 1 | Accumulation | $0.1637\tau^{0.105}$ | $0.0275\tau + 0.404$ | $0.144\tau^{0.896}$ | $1.52 - (-0.0039\tau + 0.0273)$I, $1.53 - (-0.0045\tau + 0.0262)$I, $1.54 - (-0.005\tau + 0.0252)$I, $1.54 - (-0.0059\tau + 0.0259)$I, $1.53 - (-0.0063\tau + 0.0262)$i. |
| Fine 1 | Coarse | $3.465\tau^{0.068}$ | $-0.0682\tau + 0.6968$ | $0.095\tau^{0.689}$ | |
| Fine 2 | Accumulation | $0.118 + 0.0842\tau^{0.447}$ | $0.481\tau^{0.0911}$ | $0.153\tau^{0.934}$ | $1.47 - (0.01-0.002\tau)$i, $1.48 - (0.01-0.002\tau)$i, $1.48 - (0.01-0.002\tau)$i, $1.48 - (0.01-0.002\tau)$i, $1.48 - (0.01-0.002\tau)$i. |
| Fine 2 | Coarse | $3.041 + 0.1911\tau^{2.326}$ | $0.7064 - 0.0829\tau^{0.5974}$ | $0.116\tau^{0.881}$ | |
| Fine 3 | Accumulation | $0.0376\tau + 0.1587$ | $0.4552\tau^{0.0948}$ | $0.143\tau^{0.882}$ | $1.502\tau^{0.0110} - (0.0091\tau^{0.3467})$i, $1.505\tau^{0.0123} - (0.0082\tau^{0.2866})$i, $1.506\tau^{0.0131} - (0.0074\tau^{0.2236})$i, $1.505\tau^{0.0114} - (0.0076\tau^{0.2037})$i, $1.498\tau^{0.0097} - (0.0077\tau^{0.1973})$i. |
| Fine 3 | Coarse | $0.7546\tau + 2.6882$ | $0.6256\tau^{-0.055}$ | $0.074\tau^{0.640}$ | |
| Fine 4 | Accumulation | $0.0527\tau + 0.1873$ | $0.4956\tau^{0.0948}$ | $0.1669\tau^{0.9249}$ | $(0.0088\tau + 1.4146) - (0.0066\tau^{0.2971})$i, $(0.0118\tau + 1.4145) - (0.0061\tau^{0.2398})$i, $(0.0139\tau + 1.4137) - (0.0055\tau^{0.1809})$i, $(0.0114\tau + 1.4168) - (0.0059\tau^{0.1689})$i, $(0.0085\tau + 1.4149) - (0.0060\tau^{0.1654})$i. |
| Fine 4 | Coarse | $0.3989\tau + 2.8443$ | $0.6072\tau^{-0.028}$ | $0.0818\tau^{0.8459}$ | |
| Coarse 1 | Accumulation | 0.132 | $0.5141\tau^{0.0404}$ | $0.1128\tau^{1.034}$ | $1.49\tau^{-0.017} - 0.007\tau^{-0.121}$ i, $1.50\tau^{-0.013} - 0.0055^{-0.133}$ i, $1.51\tau^{-0.01} - 0.0038^{-0.167}$ i, $1.52\tau^{-0.01} - 0.0034^{-0.3}$ i, $1.52\tau^{-0.02} - 0.0030^{-0.405}$ i. |
| Coarse 1 | Coarse | 2.542 | $0.6052\tau^{-0.059}$ | $0.5566\tau^{1.0818}$ | |
| Coarse 2 | Accumulation | $0.076 + 0.05\tau^{-0.1407}$ | $0.7357 - 0.007\tau^{-0.9275}$ | $0.0736\tau^{0.7406}$ | $1.46\tau^{-0.004} - 0.003\tau^{-0.3128}$ i, $1.48\tau^{-0.0028} - 0.0019\tau^{-0.3813}$ i, $1.49\tau^{-0.0026} - 0.0008\tau^{-0.4019}$ i, $1.48\tau^{-0.0072} - 0.0008\tau^{-0.5629}$ i, $1.47\tau^{-0.0134} - 0.0008\tau^{-0.7573}$ i. |
| Coarse 2 | Coarse | 2.172 | $0.516 - 0.0438\tau^{-0.7547}$ | $0.652\tau^{1.204}$ | |
| Continental | Water Soluble | 0.176 | 1.09 | 3.05 | 1.53 -0.005i, 1.53 – 0.005i, 1.53 – 0.005i, 1.53 – 0.005i, 1.52 – 0.017 |
| Continental | Dust | 17.6 | 1.09 | 7.364 | 1.53 – 0.008, 1.53 – 0.008, 1.53 – 0.008, 1.53 – 0.008, 1.52 – 0.008 |
| Continental | Soot | 0.050 | 0.693 | 0.105 | 1.75 – 0.46i, 1.75 – 0.45i, 1.75 – 0.45i, 1.75 – 0.45i, 1.75 – 0.44i |

Table 4. Single Scattering Albedo (SSA), Asymmetry Parameter (ASYM), and Extinction Coefficient (α) for derived and continental Aerosol Model. The optical properties tabulated here are calculated using Mie theory for AOD=0.5 at 550nm.

| Aerosol Model | SSA (λ= 0.44, **0.55**, 0.675, 0.87, 1.02 nm) $10^{-1}$ | ASYM (λ= 0.44, **0.55**, 0.675, 0.87, 1.02 nm) $10^{-1}$ | α (λ= 0.44, **0.55**, 0.675, 0.87, 1.02 nm) $10^{-2}$ |
|---|---|---|---|
| Continental | 8.92, **8.81**, 8.74, 8.59, 8.48 | 6.43, **6.39**, 6.40, 6.44, 6.53 | 7.913, **6.303**, 4.963, 3.648, 3.100 |
| Fine 1 | 8.69, **8.66**, 8.49, 8.10, 7.79 | 6.68, **6.22**, 5.74, 5.16, 5.04 | 6.658, **4.783**, 3.308, 1.993, 1.572 |
| Fine 2 | 9.37, **9.32**, 9.24, 9.07, 8.94 | 7.02, **6.66**, 6.30, 5.79, 5.61 | 8.773, **6.618**, 4.747, 2.981, 2.371 |
| Fine 3 | 9.55, **9.50**, 9.50, 9.37, 9.22 | 6.88, **6.57**, 6.15, 5.58, 5.37 | 10.08, **7.136**, 5.013, 3.036, 2.358 |
| Fine 4 | 9.62, **9.60**, 9.64, 9.47, 9.40 | 7.43, **7.10**, 6.74, 6.23, 5.98 | 12.110, **9.347**, 6.800, 4.320, 3.381 |
| Coarse 1 | 8.93, **8.97**, 9.20, 9.20, 9.23 | 6.87, **6.67**, 6.49, 6.43, 6.44 | 3.567, **2.836**, 2.325, 1.899, 1.764 |
| Coarse 2 | 9.29, **9.52**, 9.75, 9.77, 9.80 | 7.18, **6.99**, 6.82, 6.79, 6.81 | 1.079, **0.948**, 0.851, 0.759, 0.740 |

Figure 8 provides the plots of aerosol phase function at 550nm and AOD = 0.5 at 550nm for the derived AEROEX aerosol models along with the continental model, respectively. The zoomed curve at larger scattering angles is also shown. It is evident that the scattering functions of Coarse-1 and Coarse-2 models are significantly different from each other and from continental aerosols, especially for larger scattering angles. Non-absorbing Fine-3 and Fine-4 models also show significantly different scattering functions agreeing with their different asymmetry parameters. Figure 9 shows the spectral dependence of fundamental aerosol optical properties (Spectral AOD, SSA, and ASYM) of the derived AEROEX aerosol models and the continental model. Both coarse models show different spectral dependencies of SSA, ASYM, and AOD, proving that Coarse-1 and Coarse-2 are clearly different aerosol models and may have different source components. Coarse-1 shows more spectral-dependency in all three parameters, while Coarse-2 is relatively flatter. The non-absorbing and spectral flatness of optical properties implies that Coarse-2 is essentially of mineral dust origin. However, guessing the origin of Coarse-1 type aerosol based on observed optical properties (absorbing nature and spectral dependency) is not straightforward and further investigation with complimentary information is required.

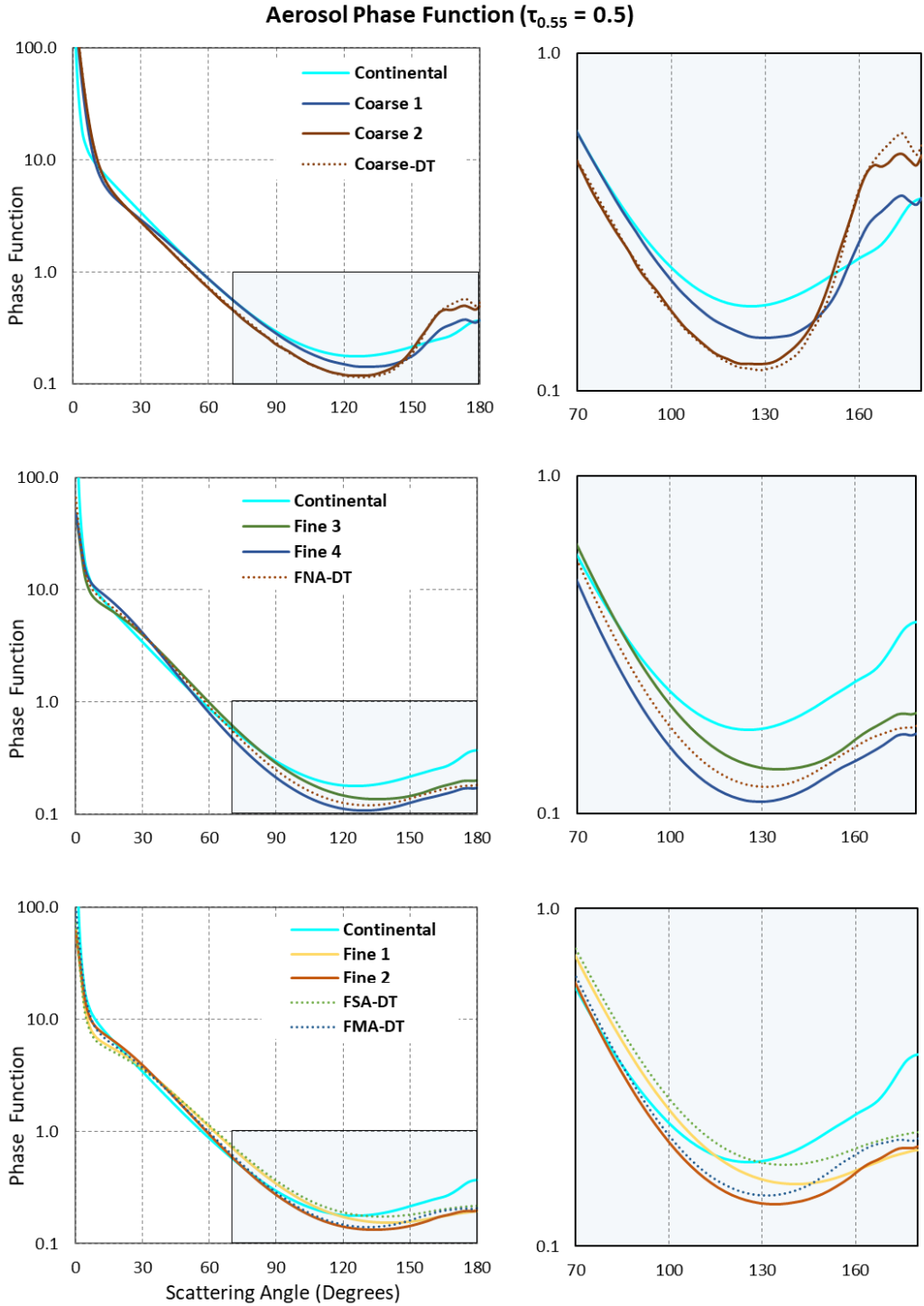

Figure 8. Aerosol Phase Function at 550 nm and AOD (550nm) = 0.5 for the Fine and coarse derived Aerosol Models and comparison with fine non-absorbing (FNA-DT), Fine Moderately Absorbing (FMA-DT), Fine Strongly Absorbing (FSA-DT) and Coarse (Coarse-DT) mode aerosol model reported in Levy et al. 2007.

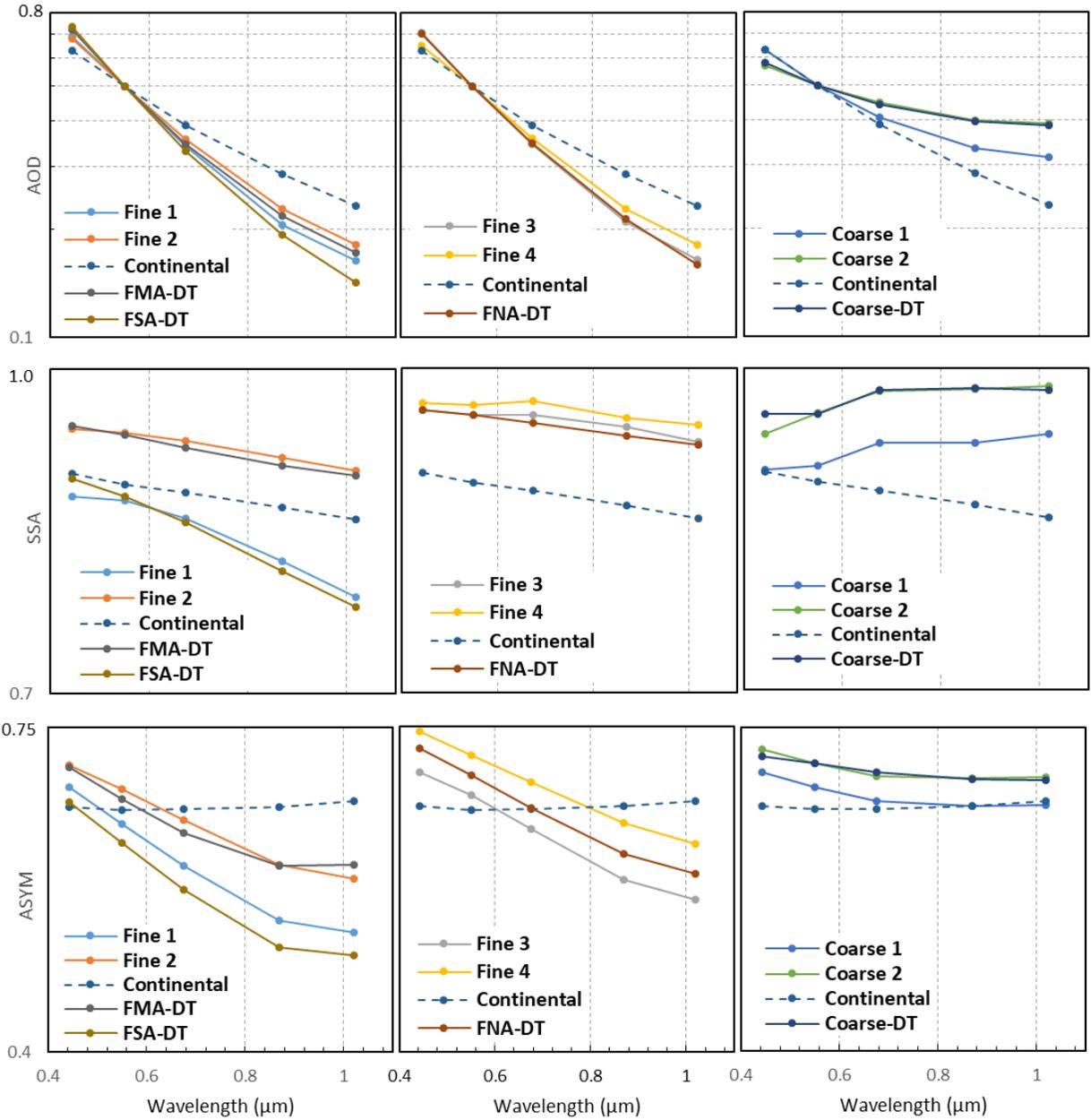

Figure 9. Spectral Dependence of Aerosol Optical Depth (AOD), Single Scattering Albedo (SSA) and Asymmetry parameter (ASYM) at AOD (550nm)=0.5 for the Fine and coarse derived Aerosol Models and comparison with fine non-absorbing (FNA-DT), Fine Moderately Absorbing (FMA-DT), Fine Strongly Absorbing (FSA-DT) and Coarse (Coarse-DT) mode aerosol model reported in Levy et al. 2007.

## 4.3 Sources of Derived Aerosol Models and Comparison with MODIS Dark-Target Models.

MODIS dark-target (DT) AOD product (both 10km and 3 km spatial resolution) is one of the most widely used and longest available AOD products at the global scale. The good accuracy of DT aerosol retrieval is primarily due to the use of dynamic mixing of radiance contributed to top-of-atmosphere (TOA) signal by fine and coarse dominated aerosol models. The currently operational DT retrieval algorithm uses fine and coarse aerosol model and the seasonal model assignment maps developed by Levy et al, (2007). Following the success of MODIS AOD product, the DT retrieval method has been made operational with many other sensors such as VIIRS, GOES-W, GOES-E and Himawari-8. MODIS (terra and aqua) AOD product has been used for finding solutions to numerous geophysical problems and is used in climate modeling as well as for data assimilation in global weather forecast models coupled with chemistry models. DT aerosol products from MODIS and subsequently from many other sensors have established the potential of satellite-based observations for aerosol remote sensing and air-quality monitoring. As discussed in the introduction, the operational DT retrieval algorithm utilizes the aerosol models developed by Levy et al 2007) using cluster analysis of 10-year AERONET data from 1993 to 2005 over around 250 locations. The four aerosol models by Levy et al., 2007 comprised three fine dominated and one coarse dominated aerosol model. For comparison of AEROEX aerosol models with those in operations with DT retrieval method, Figures 8, 9 and 10 show plots of the phase scattering function, fundamental optical properties and particle distribution function of DT models along with continental and AEROEX models. The names of DT models are as follows: Fine mode Strongly Absorbing (FSA-DT), Fine mode Moderately Absorbing (FMA-DT), Fine mode Non-Absorbing (FNA-DT), and Coarse mode dust (Coarse-DT) aerosol model. The figure 11 and 12, respectively, shows the pie charts of coarse and fine AEROEX models for each AERONET station, while figure 13 and 14, respectively, shows the most dominant coarse and fine AEROEX model found at each AERONET station location. In Figure 11-14 separate maps are generated for four quarters of the year: December-January-February (DJF), March-April-May (MAM), June-July-August (JJA) and September-October-November (SON). For northern hemisphere DJF, MAM, JJA and SON, respectively, represent winter, spring, summer and fall seasons , while for southern hemisphere the seasons are opposite. The procedure followed to generate figures 11-14 is described in next section.

Figure 8 and 9 shows that optical properties of the Coarse-DT model representing mineral dust (Levy et al., 2007) are quite similar to that of Coarse-2 aerosol model. Figure 10 shows that even microphysical properties (particle distribution function) of Coarse-2 and Coarse-DT are near to each other. From Figures 11 and 13 it is evident that the Coarse-2 models are predominantly found in the arid and semi-arid regions such as the Saharan Desert, Arabian Peninsula, Middle East, and Taklimakan Desert. This fact is in agreement with seasonal average dust-column density and wind patterns generated using MERRA-2 model reanalysis data (Figure 15). Figure 15 also shows transported dust over European countries and some parts of the North American continent, which is in agreement with the observation of the Coarse-2 aerosol model in these regions. Thus, it can be unambiguously concluded that the Coarse-2 model represents Non-Absorbing Mineral dust aerosols with deserts as source regions. In contrast to the properties of mineral dust (Coarse-2 and Coarse-DT model), the optical properties of the Coarse-1 model show an absorbing nature (Figures 8 and 9). It is interesting to see that the SSA of Coarse-1 model is even lesser than that of Fine-2 model. Interestingly the Coarse-1 aerosol model are specifically observed in the Indian subcontinent, Southeast Asia, parts of China and western Africa (see figure 11 and 13). From Figure

15 it is clear that all these regions are heavily affected by transported desert dust as well as from active fires (both wildfires as well as anthropogenic combustion activities) in different seasons. The larger mineral dust particle often acts as a sponge allowing absorption of fine carbonaceous particles on their surface leading to black carbon coat over mineral dust particles (Scarnato et al., 2015, Anderson et al., 2003). This hints that the Coarse-1 model represents carbon-coated dust particles where carbon coating gives an absorbing nature with spectrally dependent optical properties and coarser size giving the higher value of PLDR. Fine-1 model shows similar properties to that of the FSA-DT model. Fine-2 is different from both FSA-DT and FMA-DT, however more near to FMA-DT. The comparison shows that our non-absorbing fine models Fine-3 and Fine-4 possess different optical and physical properties than that of the non-absorbing model (FNA-DT) in use with the operational DT algorithm. Thus, excluding Fine-1 and Coarse-2 that are similar to Levy's models FSA-DT and Coarse-DT, respectively, rest three fine (2, 3, 4) and Coarse-1 models are new in terms of optical/physical properties. The observed differences in properties of AEROEX with DT aerosol models and need of extra fine and coarse models to represent global scenario is witnessing changes in aerosol types over the last 2-decades or inaccuracy in DT models may be due to the use of smaller AERONET dataset not capture true spatio-temporal variability at global scale.

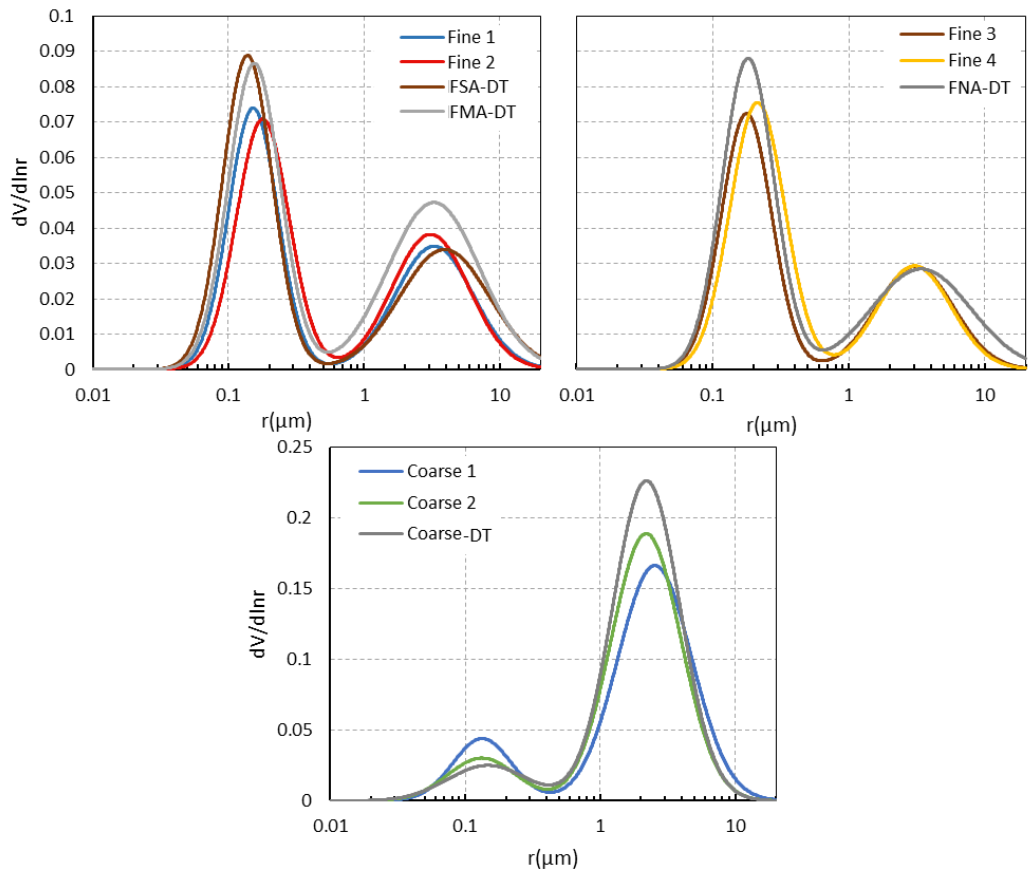

Figure 10. Comparison of Particle Size Distribution Curve for the derived Fine (Fine 1, Fine 2, Fine 3, and Fine 4) and Coarse (Coarse 1, and Coarse 2) dominated Aerosol Models with aerosol models developed by Levy et al 2007 and operational used in used in Dark target (DT) retrieval algorithm. Here FSA-DT, FMA-DT, FNA-DT and Coarse-DT refer to Fine strongly absorbing, Fine moderately adsorbing, fine non-absorbing and coarse mode aerosol models used in DT aerosol retrievals.

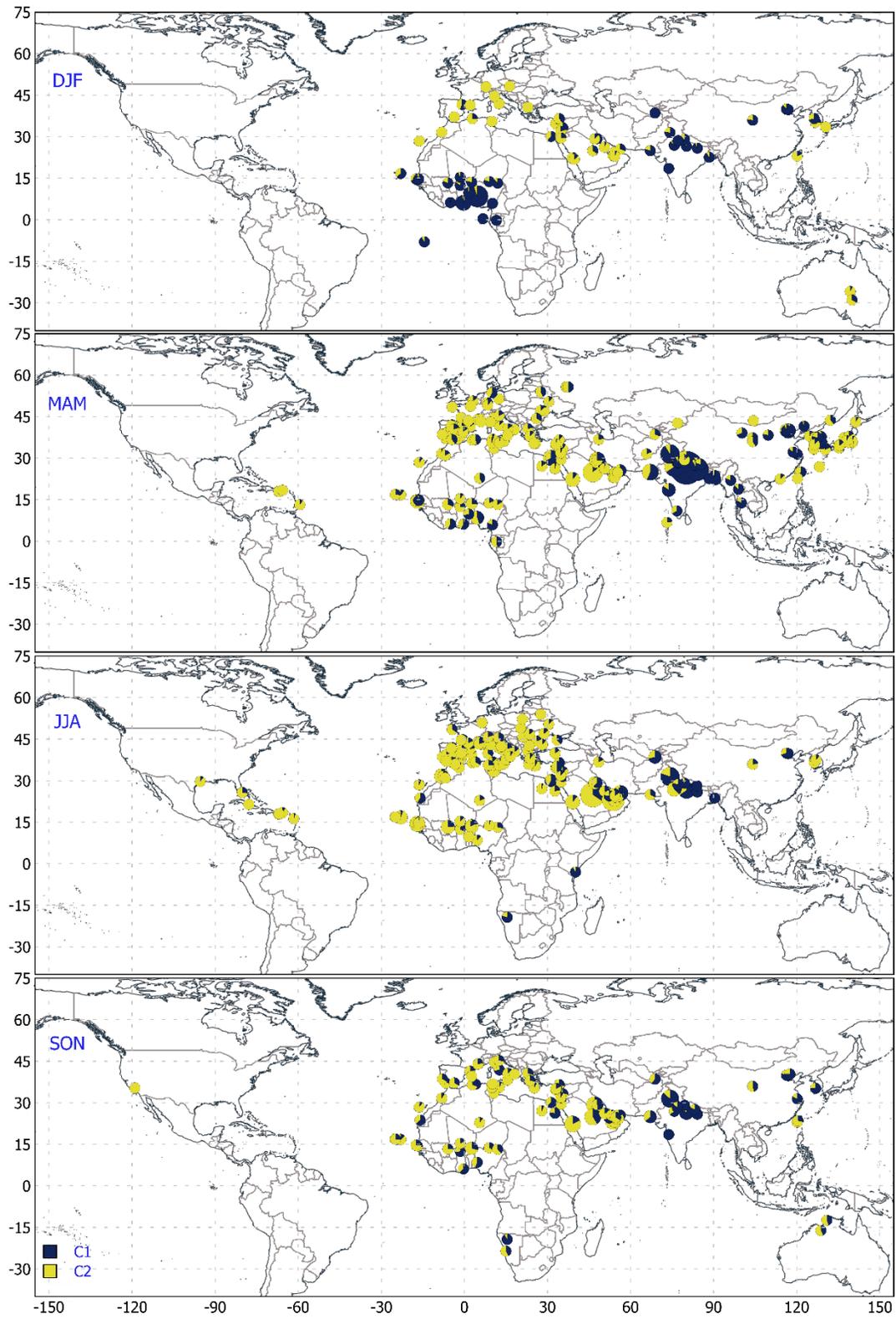

Figure 11. Pie chart representing the percentage of the derived Non – spherical Aerosol Models – Coarse 1 and Coarse 2 at each AERONET site per season.

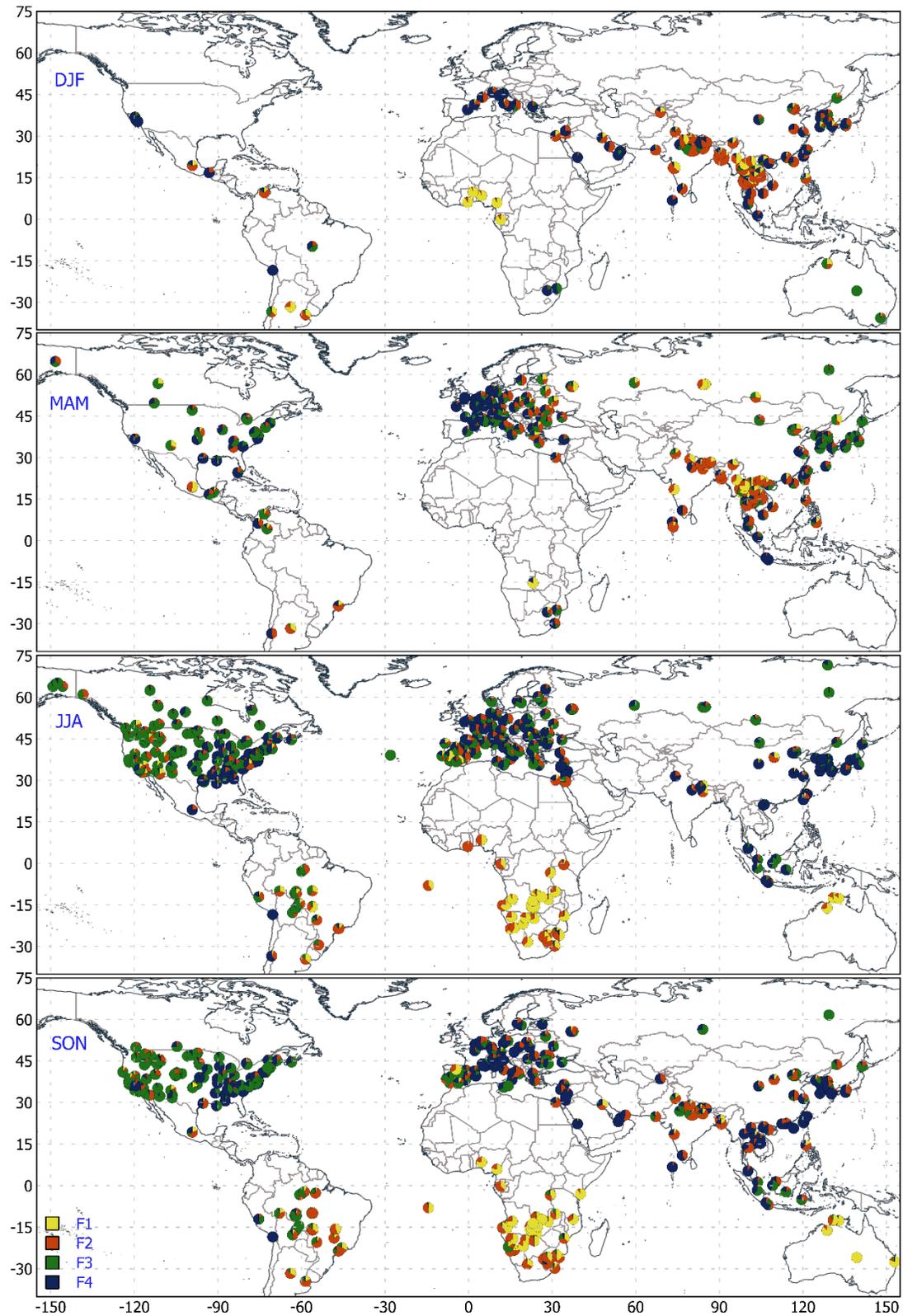

Figure 12. Pie chart representing the percentage of the derived Spherical Aerosol Models – Fine 1, Fine 2, Fine 3 and Fine 4 at each AERONET site per season.

Close investigation of Figures 12 and 15 clearly shows that Fine-1 represents the strongly absorbing carbonaceous smoke particles emitted from active fires (mostly wildfires) mainly in Western and Southern Africa, and Australia, especially during wildfire seasons of JJA, SON, and DJF. Only a few sites in South America and Southeast Asia show the Fine-1 as a prominent model during the peak wildfire season of MAM. Regions across the globe where Fine-2 aerosols are prominent such as southeast Asia, the Indian subcontinent, China, and parts of South and North America are affected by transported smoke from wildfires (Figure 14), local industrial and vehicular emissions, and other natural aerosol sources, thus Fine-2 model are mixed aerosol of continental origin with the shoot as a major component giving absorbing nature to them. Fine-2 and FMA-DT model, though have similar SSA values but have significantly different real and imaginary component of refractive index. Fine-3 and Fine-4 are non-absorbing in nature and are mainly found prominent in regions and seasons less affected by wildfires such as in North America, Western South America, European landmass, and during the season of JJA in Asia. Fine-3 and Fine-4 both are mainly of anthropogenic origin, specifically, these are sulfate and ammonium aerosols formed due to chemical reactions in the atmosphere with Sulphur dioxide and ammonia sourced from industrialization, fuel combustion, and agriculture practices. Both Fine-3 and Fine-4 show similar SSA but Fine-4 are strongly forward scattering with significantly higher ASYM value of 0.71. It is well known that sulfates and ammonia aerosols (water-soluble aerosols) are highly hygroscopic in nature. Thus phase scattering function and ASYM are highly dependent on relative humidity of the lower atmosphere. It is evident from the Figure 12 that Fine-4 aerosols are mostly prominent on the Eastern coast of North America, while Fine-3 are prominently found on the Western Coast. This difference may be attributed to high humidity over the Eastern coast leading to swelling of water-soluble aerosol in turn giving strong forward scattering properties to sulfate aerosols detected as Fine-4 aerosol model. Similarly, due to high humidity Fine-4 is more prominent in southwest Europe in comparison to northeast Europe. In India also during JJA when humidity is high and there are no wildfires the Fine-4 aerosol models becomes prominent. Thus, both Fine-3 and Fine-4 are mainly sulfates and ammonia aerosols from common sources however have different size distribution properties and forward scattering strength because of their hygroscopic nature and differences in atmospheric humidity conditions over source region. Note that excluding a few parts, the entire North America and Europe were represented by the FMA-DT model in Levy et al. (2007), whereas current study found that non-absorbing fine-3 and fine-4 can explain the aerosol types over the year in these areas. The new set of fine and coarse aerosol models are developed by using AERONET data acquired over an long period of 30 years and extended global coverage of over 900 stations, therefore we call these by name AEROEX (AERONET Extended) Aerosol Models. Based on the facts discussed above and in section 4.2, we give the following nicknames to our fine and coarse mode AEROEX aerosol models:

Fine-1 (SSA/g=0.86/0.63)    →    Strongly-Absorbing Gentle-Forward Scattering (SAGFS)

Fine-2 (SSA/g=0.93/0.67)    →    Moderate-Absorbing Moderate-Forward Scattering (MAMFS)

Fine-3 (SSA/g=0.96/0.66)    →    Non-Absorbing Moderate-Forward Scattering (NAMFS)

Fine-4 (SSA/g=0.96/0.71)    →    Non-Absorbing Strong-Forward Scattering (NASFS)

Coarse-1 (SSA/g=0.89/0.67)  →    Absorbing Carbon-Coated Dust (ACD)

Coarse-2 (SSA/g=0.95/0.69)  →    Non-Absorbing Dust (NAD)

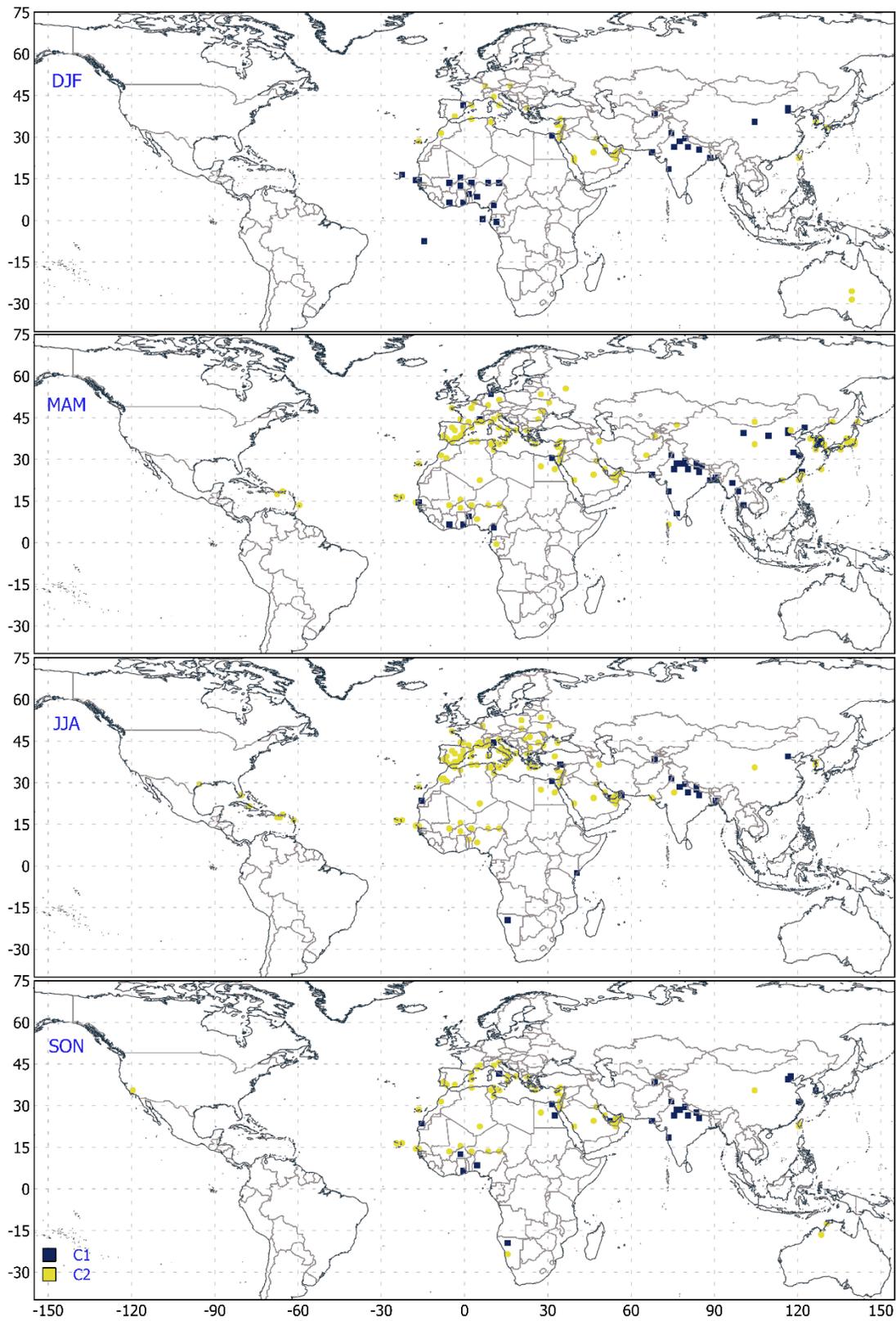

Figure 13. 1° x 1° Spatial Resolution Maps with assigned grids representing the designated Derived Non-spherical Aerosol Models – Coarse 1 and Coarse 2 per season.

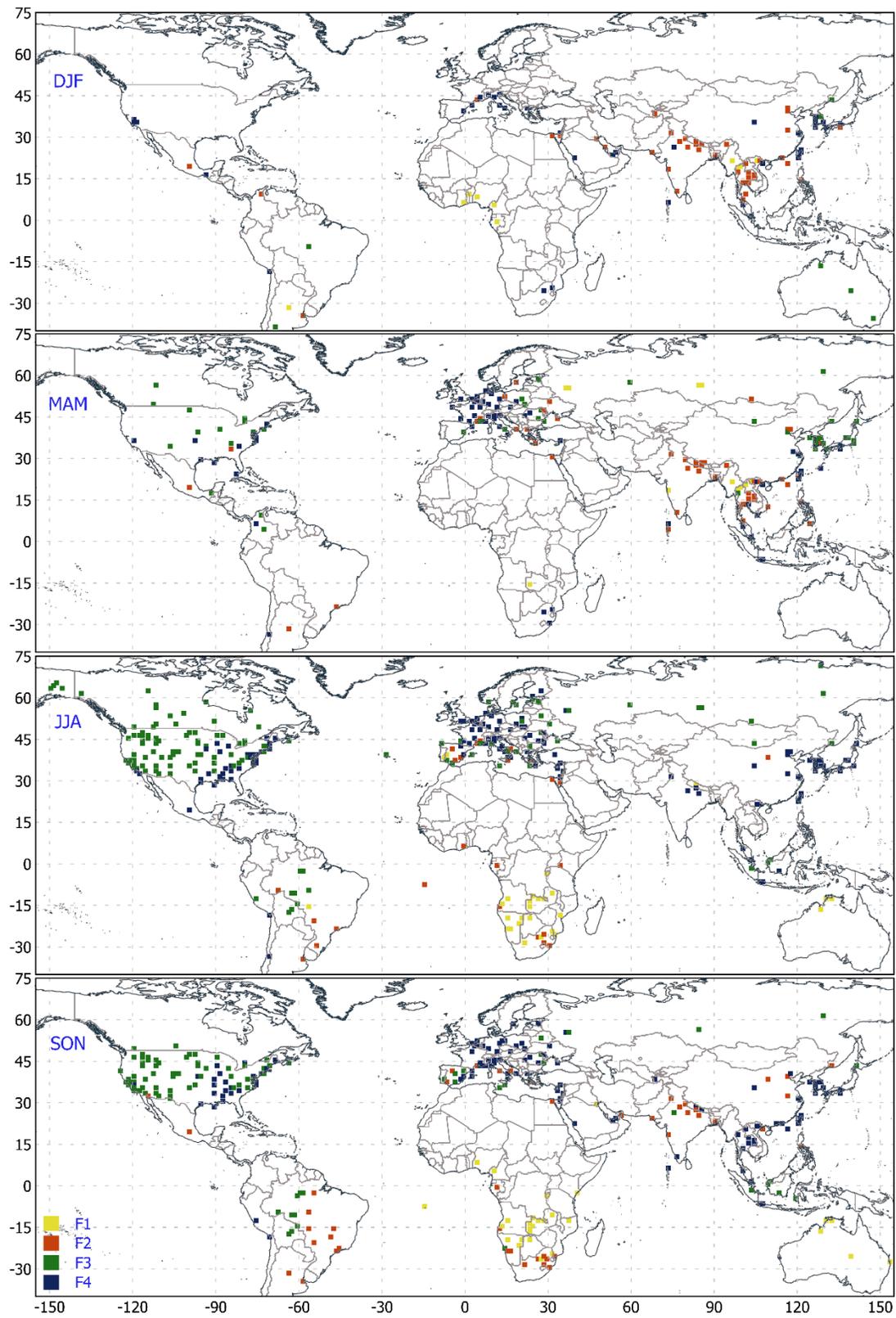

Figure 14. 1° x 1° Spatial Resolution Maps with assigned grids representing the designated Derived Spherical Aerosol Models – Fine 1, Fine 2, Fine 3 and Fine 4 per season.

## 4.4 Global Assignment of Aerosol Types

This section is focused on assigning the dominant aerosol type (fine and coarse) over land on a global map of 1-degree resolution. These assignment maps can identify the prevalent aerosol type spatiotemporally and improve the performance of satellite-based aerosol retrieval algorithms over land. Separate global maps are generated for each quarters of the year DJF, MAM, JJA SON. The initial step calculates the percentage retrievals of fine aerosol clusters for each AERONET site for all seasons.

Pie charts in Figure 11 visualize these seasonal percentages of coarse clusters for each site. Only those AERONET sites with a minimum of 10 total retrievals per season were included to ensure statistically sound results. Additionally, for each season, these sites needed at least five observations for any identified fine clusters within the minimum of 10 retrievals. This process was repeated for fine clusters, with their corresponding pie charts in Figure 12.

In the pie charts, the colored segments represent the following: yellow, red, green, and blue correspond to Fine 1, Fine 2, Fine 3, and Fine 4 of fine-dominated aerosols (Figure 12), respectively. Likewise, blue and yellow segments represent Coarse 1 and Coarse 2 of coarse-dominated aerosols (Figure 11).

Global maps with a resolution of 1 degree by 1 degree are used to assign dominant aerosol types globally for each season. The grids containing AERONET sites were assigned a dominant coarse aerosol cluster that is determined by calculating the cumulative retrieval percentage for each cluster within the grid. Seasonal plots of the assigned dominant coarse aerosol types over those grids with AERONET locations were derived, as shown in Figure 13. A similar process was applied to assign dominant fine aerosol clusters, as shown in Figure 14.

To obtain the final global assignment maps of dominant aerosol types (for both fine and coarse separately) across land areas for each season, a random forest-supervised machine learning technique was employed. The following features of the assigned grids (obtained from the previous steps) were considered for training the model: Spatial information (Latitude and longitude of the grid center), Land use (Land use data for each grid), and Coastal proximity (the grid encompasses a coastal area or not). The assigned aerosol model is used as the target variable for training.

The dominant aerosol types for all the land grids were predicted and assigned using the trained random forest model. The initial classification results are available in supplemental materials. However, these results were further refined by incorporating reasonable subjectivity to achieve the most accurate final maps. This process led to the final global assignment maps presented in Figure 16 (fine-dominated aerosols) and Figure 17 (coarse-dominated aerosols).

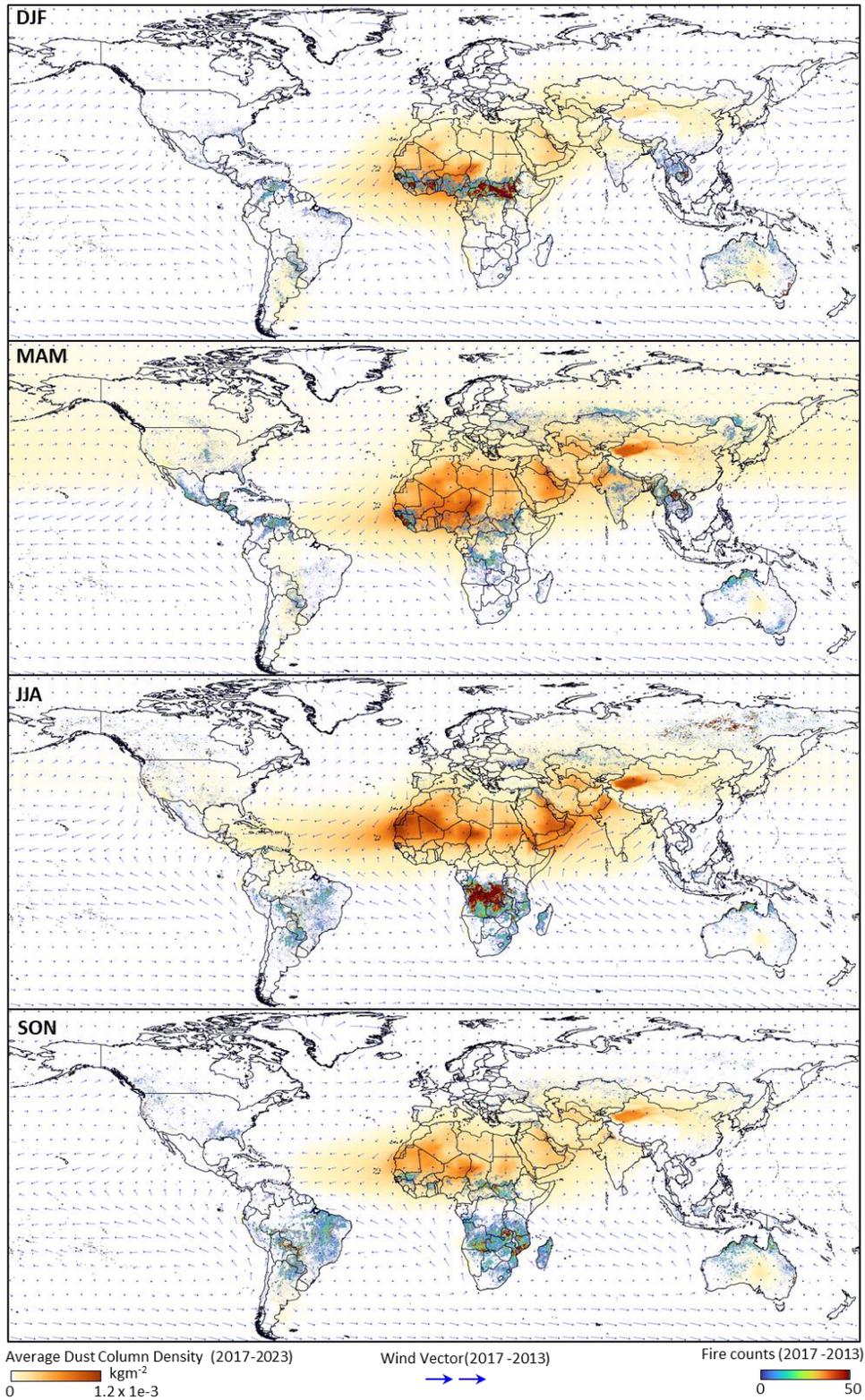

Figure 15. Seasonal average dust-column density and wind patterns generated using from MERRA-2 model reanalysis data. Also shows seasonal cumulative fire counts derived using MODIS active fire product during the period of 2017 to 2023 rasterized at spatial resolution of 10km.

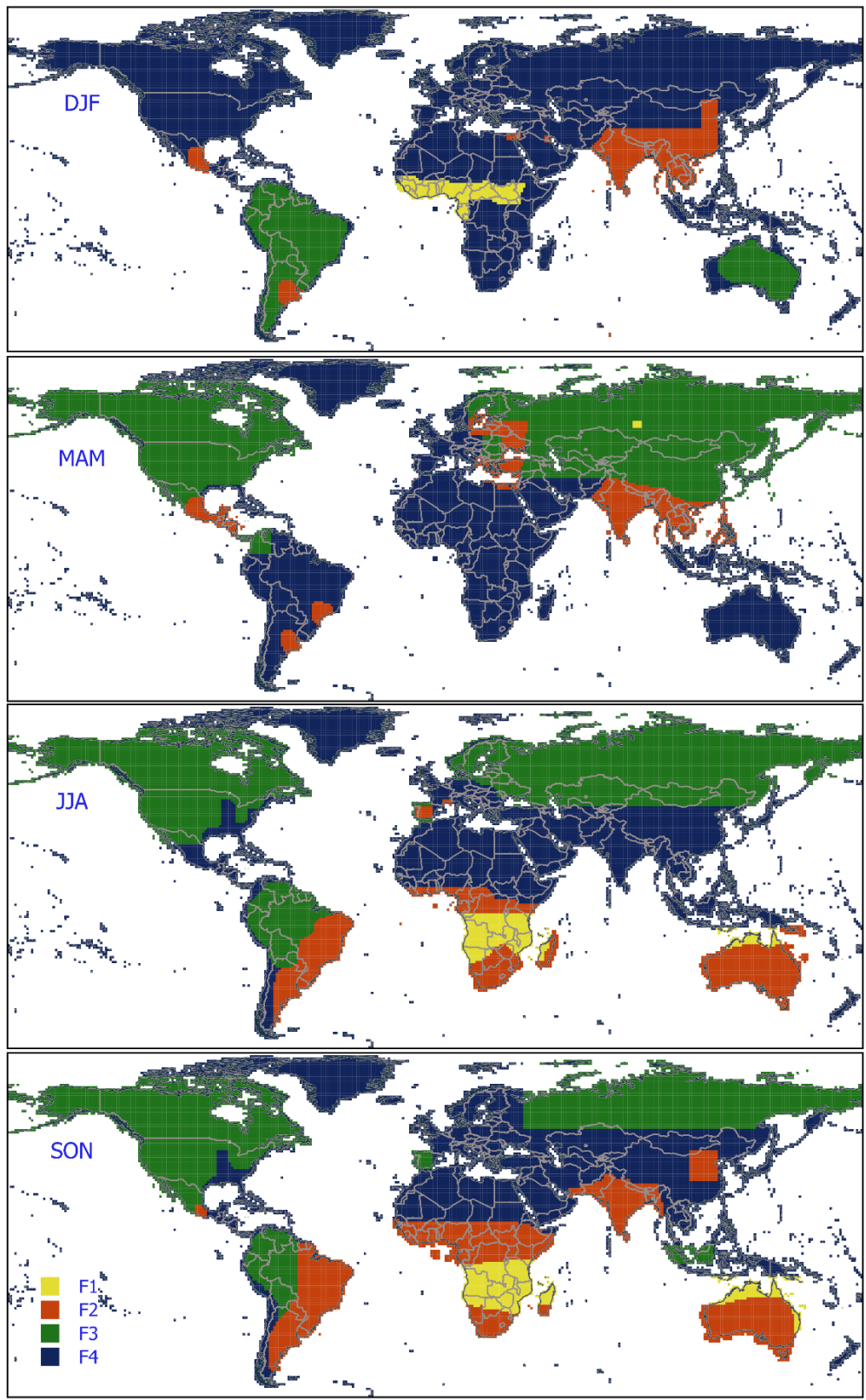

Figure 16. Global Seasonal Aerosol Assignment Map at 1° x 1° Spatial Resolution of the Derived Spherical Aerosol Models – Fine 1, Fine 2, Fine 3 and Fine 4

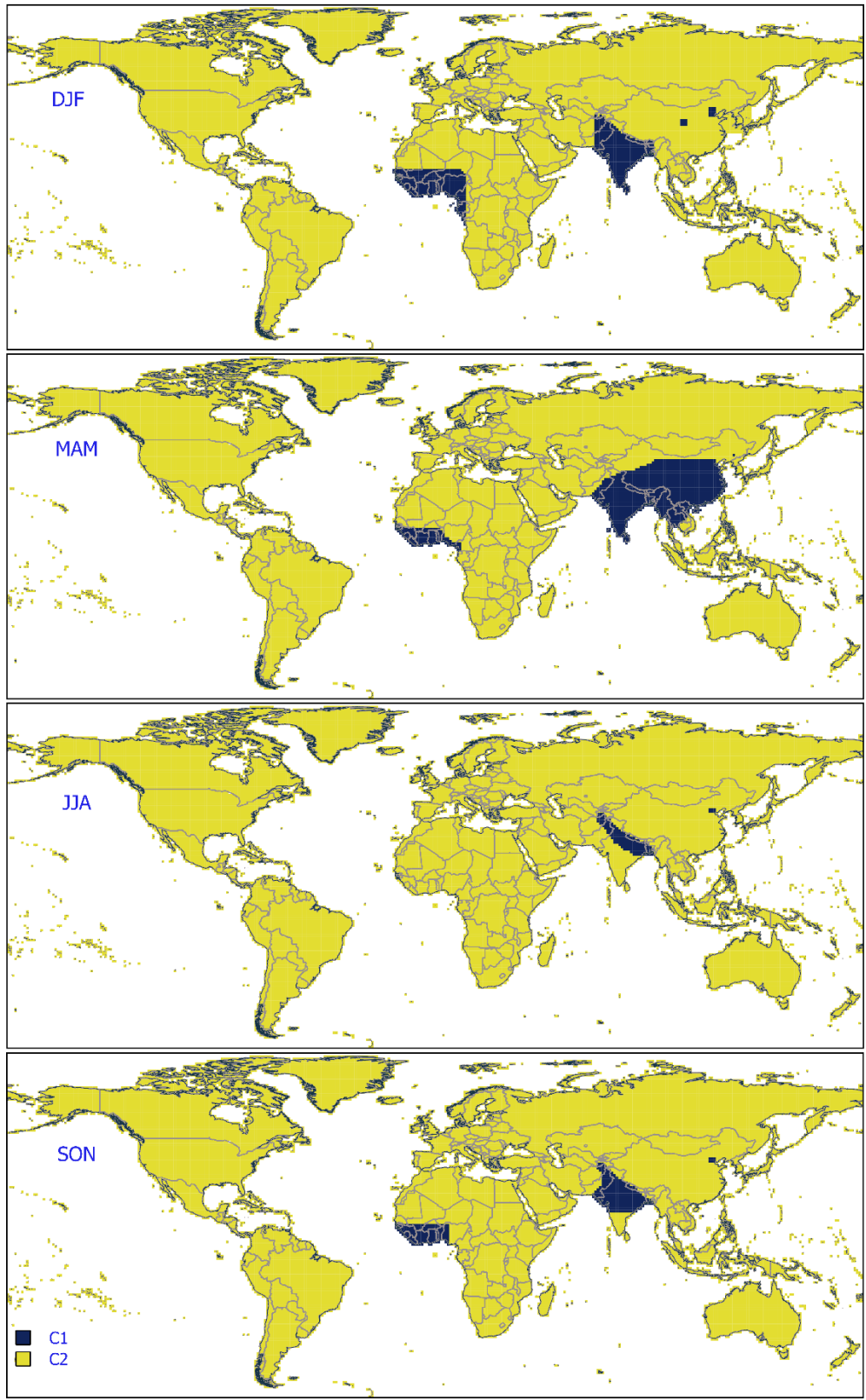

Figure 17. Global Seasonal Aerosol Assignment Map at 1° x 1° Spatial Resolution of the Derived Non–spherical Aerosol Models – Coarse 1 and Coarse 2

## 4.6 Impact on Space-Based Aerosol Remote Sensing

To analyze the implications of using AEROEX instead of DT aerosol models in satellite based aerosol optical depth retrievals, a simulation study based ground measured data from ARONET stations is performed. For simulation study, two areas are chosen, one is the Kanpur (26.51$^0$N, 80.23$^0$S) AERONET station located in Indo-Gangetic Plain of India representing complex tropical atmosphere and another is combination of three ground stations namely, Fresno_2 (36.78$^0$N, -119.77$^0$S), Missoula (46.92$^0$N,-114.08$^0$S) and Ricmrock (46.49$^0$N, -116.99$^0$S) located in North American continent, representing the Mid latitude atmosphere. The two areas chosen are of different nature in terms of aerosol sources and as well as of metrological conditions. According to Levy et al., 2007, the North American sites were represented by Moderately absorbing Fine mode (FMA-DT) aerosols, where as in current study the region is found to be affected by Non-Absorbing aerosol models NAMFS and NASFS. Similarly, for Kanpur region FSA-DT and FMA-DT were proposed by Levy et al, 2007, however current study shows presence of both Moderately as well as Non-absorbing fine aerosols represented by Models MAMFS and NASFS. For North American sites current study suggests Non-Absorbing Dust (NAD) as coarse mode aerosol which is similar to Coarse-DT model (Levy et al., 2007). However, according to current study, it is interestingly found that in almost all seasons the Indian site (Kanpur) is represented by a different coarse mode aerosol called Absorbing Carbon-Coated Dust (ACD) model. Thus keeping these facts in mind, the simulation study is performed to see implications on aerosol retrieval. The aerosol inversion products for two years 2022 and 2023 are used for simulating the 2000 top-of-atmosphere (TOA) radiances at wavelengths 440, 500, 675, 870 and 1020 nm. While performing 2000 simulation the sun-satellite geometry and aerosol sample selection is randomly generated and based on inversion product forward radiative transfer calculation are performed to generate simulated TOA radiances in mentioned wavelengths. The retrieval is then performed assuming the global maps of aerosol models operationally used in DT algorithm and using aerosol models developed in current study. Finally, comparison of AOD retrievals is performed. For estimating the uncertainty, the retrievals errors are arranged in fixed numbers of bins according to sorted AERONET AOD and for each bin 68% percentile absolute error is considered as uncertainty. The entire process is depicted in figure 18. The performance of AOD retrievals using DT and AEROEX aerosol model is shown in figure 19. For Kanpur region Figure (a) and (b) shows significantly improved AOD retrievals in terms of reduced bias, mean absolute error and root mean square error when AEROEX models are used. Similar observation sustains for USA sites. The histogram shows that peak retrieval errors are near zero for the case of retrievals with MS models while for DT models the peak error is significantly displaced from zero. Another important observation is the distribution curve is narrower about zero for the case of AEROEX models while in case of DT, the error spread is wider. Figure 20 shows the uncertainty curves for retrieved AOD as function of AERONET for the two simulation cases. In both cases the uncertainty in retrieved AOD is found to be strongly dependent on true AOD value. For tropical complex atmosphere represented by Kanpur station t both relative as well as absolute errors are significantly less when AEROEX model is used instead of DT models. Similar observation observed for North American case study but main improvement is observed in absolute error. This shows that introduction of AEROEX models in operational DT retrieval algorithm may significantly improve the MODIS AOD retrievals. In addition new and more number of fine and coarse aerosol models will surely contribute in reducing the uncertainty of aerosol radiative forcing and solving several climate related geo-physical problems.

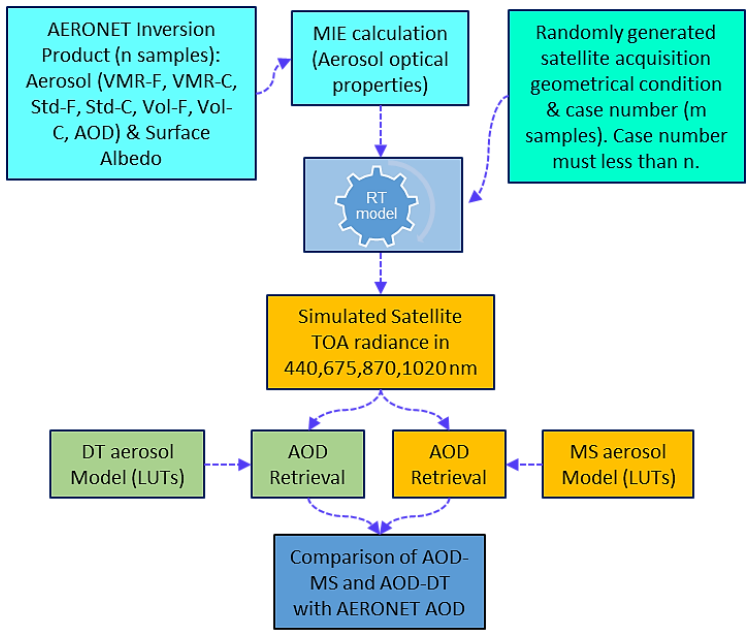

Figure 18. Block diagram showing steps involved to study the implications of derived aerosol models on satellite based aerosol retrievals in comparison to when DT aerosol models are used.

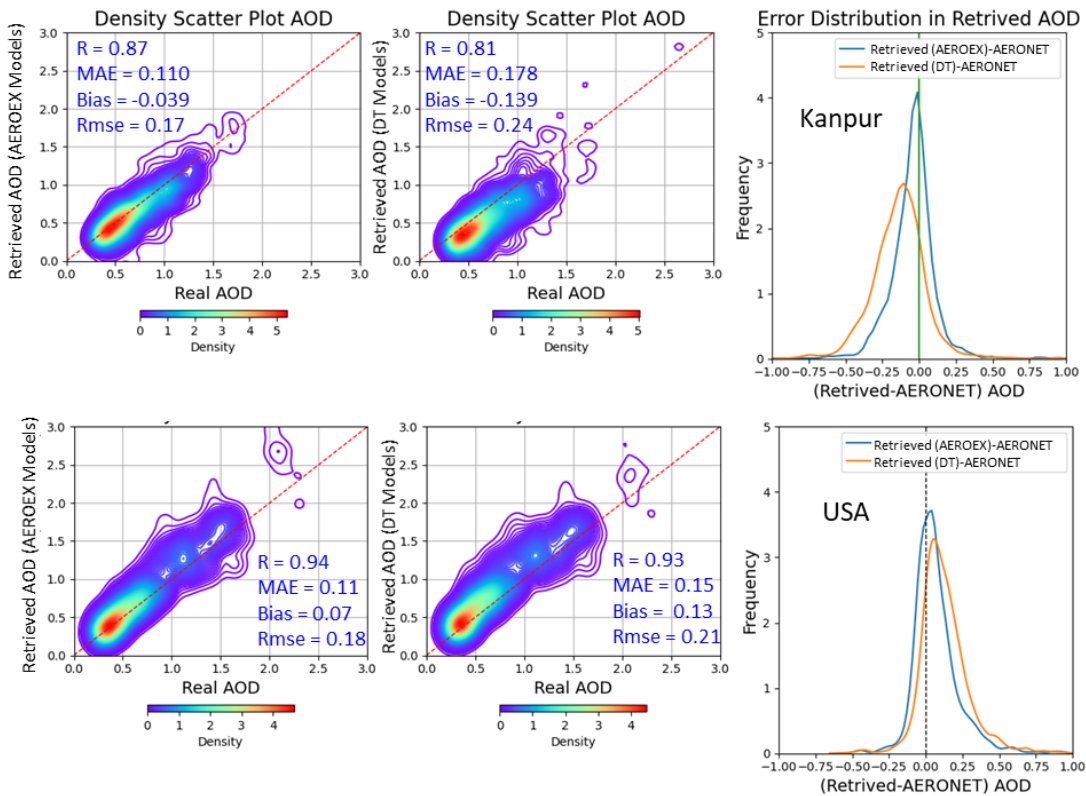

Figure 19. The first rows shows simulation results for Kanpur station. The density scatter plots in first row shows comparison of retrieved AOD with real AOD when AEROEX and DT aerosol models are used for retrieval. The distribution plot shows the error distribution curves for retrievals using AEROEX (green) and DT (red) aerosol models. Second rows shows same variables but for USA stations.

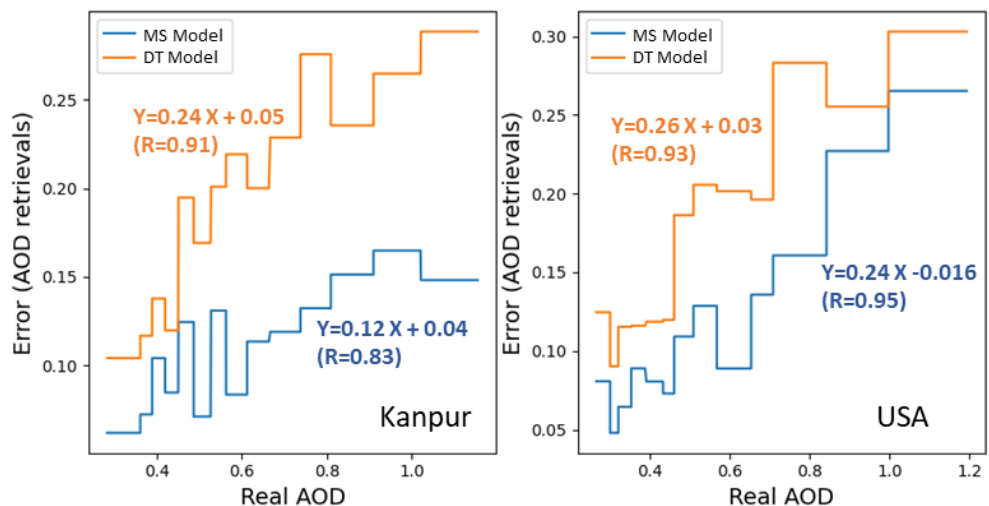

Figure 20. Left plot shows the uncertainty in AOD retrievals as a function of real AOD using AERONEX and DT aerosol models for Kanpur stations, respectively. Right plot is same as left but for USA stations.

## 5. Conclusion

The aerosol models currently used in the Dark-Target aerosol algorithm for satellite-based aerosol retrievals, such as those from MODIS and VIIRS, were originally developed by Levy et al. (2007) based on a limited set of AERONET data. While these models have been instrumental in aerosol research and have addressed various geophysical problems, the AERONET network has significantly expanded over the last two decades, now covering more than 900 stations globally. This expanded dataset, spanning nearly 30 years of aerosol data, provides a much richer foundation for aerosol modeling.

In this study, we conducted a comprehensive machine learning-based cluster analysis of aerosol data from AERONET, spanning three decades and over 900 global sites. From this, we developed a new set of fine and coarse mode aerosol models, named AEROEX, which incorporates updated microphysical parameters and optical properties. Notably, we found that four fine-mode aerosol models, rather than the three proposed by Levy et al. (2007), are necessary to accurately represent the global distribution of fine-mode aerosols.

Additionally, our analysis reveals important regional differences in aerosol types. In Europe and North America, where Levy's models previously used a moderately absorbing aerosol, we identified two distinct types of non-absorbing aerosols—moderately forward scattering and strongly forward scattering—primarily composed of sulfate. These aerosols are distinguished by their hygroscopic properties and regional/seasonal variations in humidity, which were not fully captured in earlier models.

For coarse-mode aerosols, we found that two models are necessary to accurately represent the global distribution. One of these models is similar to Levy's mineral dust model, while the other is a newly identified carbon-coated mineral dust model, which exhibits significantly different optical properties. This new understanding of coarse-mode aerosol composition, particularly in regions like the Indian subcontinent and Western Africa, offers a more accurate representation of aerosol types in these areas.

The updated global distribution maps generated from this study using the AEROEX aerosol models show significant differences from those used in the current operational Dark-Target aerosol retrieval algorithm. These revised models, with their improved accuracy, will enhance aerosol remote sensing and retrieval algorithms, reducing uncertainties in aerosol optical depth (AOD) measurements and aerosol radiative forcing estimates.

As future satellite missions, such as the ISRO-French TRISHNA, require more region-specific and seasonal aerosol models, the results of this study provide a solid foundation for improving aerosol retrievals in these upcoming missions. The AEROEX models and distribution maps developed here will be essential for advancing aerosol remote sensing techniques and improving our understanding of global aerosol dynamics.

**Acknowledgement**

Special thanks to Principal Investigators of AERONET sites across the globe for making data available through the NASA Aerosol Robotic Network (AERONET) website (https://aeronet.gsfc.nasa.gov). The authors would like to thank Mr. Deepak Putrevu, Mr. C Patnaik, Dr. Rashmi Sharma and Shri Nilesh M Desai for their support.

**Declaration**

Although this work is internally reviewed, the views and opinions expressed in this paper are those of the author and should not be interpreted as an official Indian Space Research Organisation (ISRO) or Government of India position, policy, or decision.

The author declare no conflict of interest.

**Data Availability**

AERONET data (inversion product) is freely downloadable through the NASA Aerosol Robotic Network (AERONET) website (https://aeronet.gsfc.nasa.gov).

The processed AERONET data and all derived product presented in this manuscript are publically available as a QGIS project and can be downloaded from author's GitHub repository. The QGIS project will allow readers to have an interactive experience of results presented in the paper and attribute tables will show the real data generated after processing and clustering analysis.

The MODIS fire data is freely downloadable from the Fire Information for Resource Management System (FIRMS) website https://firms.modaps.eosdis.nasa.gov.

The MERRA-2 model data used here is free available from NASA's Giovanni web portal.

**References.**

Albrecht, B. A. (1989). Aerosols, cloud microphysics, and fractional cloudiness. Science, 245(4923), 1227-1230.

Anderson, T L., S.J. Masonis, D.S. Covert, N.C. Ahlquist, S G. Howell, A.D. Clarke, C S. McNaughton, 2003. Variability of aerosol optical properties derived from in-situ aircraft measurements during ACE-Asia. Journal of Geophysical Research-Atmospheres 108 D23. https://doi.org/10.1029/2002JD003247

Che, H., X. Xia, J. Zhu, H. Wang, Y. Wang, J. Sun, X. Zhang, G. Shi, Aerosol optical properties under the condition of heavy haze over an urban site of Beijing, China, Environ. Sci. Pollut. Res., 22 (2015), pp. 1043-1053.

Dubovik, O., B.N. Holben, T. Lapyonok, A. Sinyuk, M.I. Mishchenko, P. Yang, and I. Slutsker, 2002a: Non-spherical aerosol retrieval method employing light scattering by spheroids. Geophys. Res. Lett., 29, no. 10, 1415, doi:10.1029/2001GL014506.

Dubovik, O., Brent Holben, Thomas F. Eck, Alexander Smirnov, Yoram J. Kaufman, Michael D. King, Didier Tanré, and Ilya Slutsker, 2002b, Variability of Absorption and Optical Properties of Key Aerosol Types Observed in Worldwide Locations, Journal of the Atmospheric Sciences, 59, 3, 590–608.

Eck, T., Holben, B.N., Reid, J., Dubovik, O., Smirnov, A., O'Neill, N., Slutsker, I., Kinne, S., 1999. Wavelength dependence of the optical depth of biomass burning, urban, and desert dust aerosols, J. Geophys. Res. 104(D24), 31, 333– 31,349.

Ehlers, K. M., Moosmüller, H. (2023). Small and Large Particle Limits of the Asymmetry Parameter for Homogeneous, Spherical Particles, Aerosol Sci. Technol., 57 (5), 425-433, 10.1080/02786826.2023.2186214

Fan, Y., X. Sun, H. Huang, R. Ti, X. Liu, The primary aerosol models and distribution characteristics over China based on the AERONET data, J. Quant. Spectrosc. Ra., 275 (2021), doi: 10.1016/j.jqsrt.2021.107888

Giles, D.M., B.N. Holben, T.F. Eck, A. Sinyuk, A. Smirnov, I. Slutsker, R.R. Dickerson, A.M. Thompson, J.S. Schafer, An analysis of AERONET aerosol absorption properties and classifications representative of aerosol source regions, J. Geophys. Res. Atmos., 117 (2012), pp. 1-16

Global Modeling and Assimilation Office (GMAO). 2015-06-30. M2TMNXAER. Version 5.12.4. MERRA-2 tavgM_2d_aer_Nx: 2d,Monthly mean,Time-averaged,Single-Level,Assimilation,Aerosol Diagnostics V5.12.4. Greenbelt, MD, USA. Archived by National Aeronautics and Space Administration, U.S. Government, Goddard Earth Sciences Data and Information Services Center (GES DISC). https://doi.org/10.5067/FH9A0MLJPC7N. https://disc.gsfc.nasa.gov/datacollection/M2TMNXAER_5.12.4.html. Digital Science Data.

Global Modeling and Assimilation Office (GMAO). 2015-06-30. M2TMNXFLX. Version 5.12.4. MERRA-2 tavgM_2d_flx_Nx: 2d,Monthly mean,Time-Averaged,Single-Level,Assimilation,Surface Flux Diagnostics V5.12.4. Greenbelt, MD, USA. Archived by National Aeronautics and Space Administration, U.S.

Government, Goddard Earth Sciences Data and Information Services Center (GES DISC). https://doi.org/10.5067/0JRLVL8YV2Y4. https://disc.gsfc.nasa.gov/datacollection/M2TMNXFLX_5.12.4.html. Digital Science Data.

Holben, B.N., Eck, T.F., Slutsker, I., Tanre, D., Buis, J.P., Setzer, A., Vermote, E., Reagan, J.A., Kaufman, Y.J., Nakajima, T., Lavenu, F., Jankowiak, I., Smirnov, A., 1998. AERONET-A federated instrument network and data archive for aerosol characterization. Remote sensing of environment 44 (1), 1-16.

IPCC, 2021: Climate Change 2021: The Physical Science Basis. Contribution of Working Group I to the Sixth Assessment Report of the Intergovernmental Panel on Climate Change[Masson-Delmotte, V., P. Zhai, A. Pirani, S.L. Connors, C. Péan, S. Berger, N. Caud, Y. Chen, L. Goldfarb, M.I. Gomis, M. Huang, K. Leitzell, E. Lonnoy, J.B.R. Matthews, T.K. Maycock, T. Waterfield, O. Yelekçi, R. Yu, and B. Zhou (eds.)]. Cambridge University Press, Cambridge, United Kingdom and New York, NY, USA, In press, doi:10.1017/9781009157896.

Kondragunta, S. et al., 2023. Monitoring of Surface PM2.5: An International Constellation Approach to Enhancing the Role of Satellite Observations, https://doi.org/10.25923/7snz-vn34

Lu, F., Xu, D., Cheng, Y., Dong, S., Guo, C., Jiang, X., Zheng, X., 2015. Systematic review and meta-analysis of the adverse health effects of ambient PM2.5 and PM10 pollution in the Chinese population. Environ. Res. 136, 196–204, doi:101016/j.envres.2014.06.029.

Levy, R. C., Remer, L. A., Dubovik, O., 2007. Global aerosol optical properties and application to Moderate Resolution Imaging Spectroradiometer aerosol retrieval over land. J. Geophys. Res. 112, D13210, doi:10.1029/2006JD007815.

Lee, J., J. Kim, C.H. Song, S.B. Kim, Y. Chun, B.J. Sohn, B.N. Holben, Characteristics of aerosol types from AERONET sunphotometer measurements, Atmospheric Environment, Volume 44, Issue 26, 2010, Pages 3110-3117, ISSN 1352-2310, https://doi.org/10.1016/j.atmosenv.2010.05.035.

Li, J., Carlson, B. E., and Lacis, A. A.: Using single-scattering albedo spectral curvature to characterize East Asian aerosol mixtures, J. Geophys. Res.-Atmos., 120, 2037–2052, 2015.

Logothetis, S-A., Vasileios Salamalikis, Andreas Kazantzidis, The impact of different aerosol properties and types on direct aerosol radiative forcing and efficiency using AERONET version 3, Atmospheric Research, Volume 250, 2021, 105343, ISSN 0169-8095, ttps://doi.org/10.1016/j.atmosres.2020.105343.

Louis Giglio, Jacques Descloitres, Christopher O. Justice, Yoram J. Kaufman, An Enhanced Contextual Fire Detection Algorithm for MODIS, Remote Sensing of Environment, Volume 87, Issues 2–3, 2003, Pages 273-282, ISSN 0034-4257, https://doi.org/10.1016/S0034-4257(03)00184-6.

Lenoble, J., Brogniez, C. (1984). A comparative review of radiation aerosol models. Beitr. Phys. Atmosph. 57, 1, 1-20.

Li, C.; Li, J.; Dubovik, O.; Zeng, Z.-C.; Yung, Y.L. Impact of Aerosol Vertical Distribution on Aerosol Optical Depth Retrieval from Passive Satellite Sensors. Remote Sens. 2020, 12, 1524. https://doi.org/10.3390/rs12091524

Mukherjee, A. Agrawal, M., 2017. World air particulate matter: sources, distribution and health effects. Environ. Chem. Lett. 15(2), 283–309.

Mahowald, N. M., et al. (2005). Impact of desert dust on ocean fertilization. Journal of Geophysical Research: Biogeosciences, 110(G2).

Mishra, M. K., 2018. Retrieval of aerosol optical depth from INSAT-3D imager over Asian landmass and adjoining ocean: Retrieval uncertainty and validation. J. Geophys. Res. Atmos. 123, 5484– 5508. https://doi.org/10.1029/2017JD028116.

Mishra, M. K., P.S. Rathore, A. Misra, R. Kumar. 2020. Atmospheric correction of multi-spectral VNIR remote sensing data: Algorithm and Inter-sensor comparison of aerosol and surface reflectance products. Earth and Space Science, e2019EA000710.

Mishra, M. K., A. Misra, R. Kumar. 2023. Operational AOD retrieval at subkilometer resolution using OceanSat-2 OCM over land: SAER algorithm, uncertainties, validation & inter-sensor comparison. Earth and Space Science. doi: essopenarchive.org/doi/full/10.1002/essoar.105121161.

Mishra, M. K., P.S. Rathore. 2021. Impact of nationwide Covid-19 lockdown on Indian air quality in terms of aerosols as observed from space. Aerosol and Air Quality Research 21 (4), 200461.

Noh, Y., Müller, D., Lee, K., Kim, K., Lee, K., Shimizu, A., Sano, I., and Park, C. B.: Depolarization ratios retrieved by AERONET sun–sky radiometer data and comparison to depolarization ratios measured with lidar, Atmos. Chem. Phys., 17, 6271–6290, https://doi.org/10.5194/acp-17-6271-2017, 2017.

OECD (2016). The Economic Consequences of Outdoor Air Pollution. OECD Publishing.

Omar, A. H., J.-G. Won, D. M. Winker, S.-C. Yoon, O. Dubovik, and M. P. McCormick (2005), Development of global aerosol models using cluster analysis of Aerosol Robotic Network (AERONET) measurements, J. Geophys. Res., 110, D10S14, doi:10.1029/2004JD004874.

Qin, Y., and R. M. Mitchell (2009), Characterisation of episodic aerosol types over the Australian continent, Atmos. Chem. Phys., 9, 1943–1956, doi:10.5194/acp-9-1943-2009.

Russell, P., R. Bergstrom, Y. Shinozuka, A. Clarke, P. DeCarlo, J. Jimenez, J. Livingston, J. Redemann, O. Dubovik, A. Strawa Absorption Angstrom Exponent in AERONET and related data as an indicator of aerosol composition, Atmos. Chem. Phys., 10 (2010), pp. 1155-1169Giles et al., 2012

Remer, L.A., Kaufman, Y.J., Tanre, D., Mattoo, S., Chu, D.A., Martins, J.V., Li, R.R., Ichoku, C., Levy, R.C., Kleidman, R.G., Eck, T.F., Vermote, E., Holben, B.N., 2005. The MODIS aerosol algorithm, products, and validation. J. Atmos. Sci., 62, 947–973, https://doi.org/10.1175/JAS3385.1.

Remer, L.A., Mattoo, S., Levy, R.C., Munchak, L.A., 2013. MODIS 3 km aerosol product: algorithm and global perspective. Atmospheric Measurement Tech. Discuss. 6, 69-112.

Shin, S.-K., Tesche, M., Noh, Y., and Müller, D.: Aerosol-type classification based on AERONET version 3 inversion products, Atmos. Meas. Tech., 12, 3789–3803, https://doi.org/10.5194/amt-12-3789-2019, 2019.

Scarnato, B. V., S China, K. Nielsen, C. Mazzoleni, 2015. Perturbations of the optical properties of mineral dust particles by mixing with black carbon: A numerical simulation study. Atmospheric Chemistry and Physics Discussion 15, 2487-2533. https://doi.org/10.5194/acpd-15-2487-2015.

Twomey, S. (1977). The influence of pollution on the shortwave albedo of clouds. Journal of the Atmospheric Sciences, 34(7), 1149-1152.

Velasco-Merino, C., Mateos, D., Toledano, C., Prospero, J.M., Molinie, J., Euphrasie-Clotilde, L., González, R., Cachorro, V.E., Calle, A., and de Frutos, A.M., 2018. Impact of long-range transport over the Atlantic Ocean on Saharan dust optical and microphysical properties based on AERONET data. Atmos. Chem. Phys. 18, 9411–9424, https://doi.org/10.5194/acp-18-9411-2018.

Vernier, J.-P., Fairlie, T., Natarajan, M., Deshler, T., Gadhavi, H., Ratnam, M., et al. (2018). Batal: The balloon measurement campaigns of the Asian tropopause aerosol layer. Bulletin of the American Meteorological Society, 99(5), 955–973. https://doi.org/10.1175/BAMS-D-17-0014.1

Wu, X., Griessbach, S., Hoffmann, L., 2018. Long-range transport of volcanic aerosol from the 2010 Merapi tropical eruption to Antarctica. Atmos. Chem. Phys. 18, 15859–15877, https://doi.org/10.5194/acp-18-15859-2018.

Zhang, Z. Wang, J., Chen, L., Chen, X., Sun, G., Zhong, N., Kan, H., Lu, W., 2014a. Impact of haze and air pollution-related hazards on hospital admissions in Guangzhou. China. Environ. Sci. Pollut. Res. 21, 4236–4244.

Zhang, Z., et al. (2014b). Fine particulate matter (PM2.5) and respiratory tract development: A longitudinal cohort study in Chinese children. Journal of Thoracic Disease, 6(7), 850-861.

Zhang, L., J. Li, Variability of major aerosol types in China classified using AERONET measurements, Remote Sens., 11 (20) (2019), 2334; https://doi.org/10.3390/rs11202334